# Process Algebra as Abstract Data Types *

**Ruqian Lu[1,2,3,4,§], Lixing Li[1,2], Yun Shang[1,2], Xiaoyu Li[5]**

[1]Institute of Mathematics, Academy of Mathematics and Systems Science, CAS
[2]CAS Key Lab of MADIS, Academy of Mathematics and Systems Science, CAS
[3]Shanghai Key Lab of Intelligent Information Processing, Fudan University,
[4]CAS Key Lab of IIP, Institute of Computing Technology, CAS
[5]School of Information Engineering, Zhengzhou University

* Partially supported by NSFC project 60603002 & 60736011, 973 project 2009CB320701, 863project 2007AA01Z325,
§ Corresponding author: rqlu@math.ac.cn. Other authors: Lixing Li: llx0331@163.com, Yun Shang: shangyun602@163.com, Xiaoyu Li: iexyli@zzu.edu.cn

# Abstract

In this paper we introduced an algebraic semantics for process algebra in form of abstract data types. For that purpose, we developed a particular type of $\Sigma$ algebra, the seed algebra, which describes exactly the behavior of a process within a labeled transition system. We have shown the possibility of characterizing the bisimulation of two processes with the isomorphism of their corresponding seed algebras. We pointed out that the traditional concept of isomorphism of $\Sigma$ algebra does not apply here, because there is even no one-one correspondence between the elements of two seed algebras. The lack of this one-one correspondence comes from the non-deterministic choice of transitions of a process. We introduce a technique of hidden operations to mask unwanted details of elements of a seed algebra, which only reflect non-determinism or other implicit control mechanism of process transition. Elements of a seed algebra are considered as indistinguishable if they show the same behavior after these unwanted details are masked. Each class of indistinguishable elements is called a non-hidden closure. We proved that bisimulation of two processes is equivalent to isomorphism of non-hidden closures of two seed algebras representing these two processes. We call this kind of isomorphism a deep isomorphism. We get different models of seed algebra by specifying different axiom systems for the same signature. Each model corresponds to a different kind of bisimulation. By proving the relations between these models we also established relations between 10 different bisimulations, which form a acyclic directed graph.

## $ 1 Introduction

### 1.1 Process algebras and their formal semantics

The year 1971 can be considered as the birth year of concept of process algebra. Hans Beki$\hat{c}$ has proposed the concept and a prototype of process algebra in his seminal paper 'Towards a mathematical theory of processes'[9]. He said in this paper that 'our plan to develop an algebra of processes may be viewed as a high-level approach: we are interested in how to compose complex processes from simpler (still arbitrary complex) ones'. The first process algebra Communicating Sequential Processes (CSP) was proposed in 1978 by Tony Hoare[30]. It is of an imperative style. Another process algebra of functional style, Calculus of Communicating Systems (CCS), was proposed by Robin Milner[39]. Although Milner's research on this topic had started many years ago, CCS in its perfect form was published in 1989 in form of a book[40]. The third prominent early process algebra, Algebra of Communicating Processes (ACP), was proposed first without communication (called PA) by Jan Bergstra and Jan Willem Klop[11], then in its perfect form (with communication) in 1984[12]. Since that time the research on process algebra has produced proliferent results. To name a few, there are process algebras for mobile communication ($\pi$ calculus)[41], for site sensitive communication (Ambiant)[20], for secure communication (Spi)[4], for performance evaluation (PEPA)[25], etc. The semanrtics of these process algebras has been a very hot topic of research that has produced lots of papers.

In this paper, we are not interested in the big variety of the semantics of these different process algebras. Each of them has its own semantics. What concerns us is the common properties of different semantic approaches. No matter how different their semantics are, the approaches they accept to define their semantics are more and less similar and are limited to a small number of styles. The mostly often used semantic approach is the structural operational approach invented by G. Plotkin[43]. Almost all important programming languages have adopted structural operational approach to define their semantics since that time. Process algebra is not an exception. With this approach, the operational semantics of a process algebra is described by a labeled transition system (LTS). The transition rules look like

$$P \xrightarrow{\alpha} P' \qquad \text{or} \qquad \frac{P \xrightarrow{\alpha} P' \& other..conditions}{Q \xrightarrow{\beta} Q'}$$

Where the rule on the left means the process P performs action $\alpha$ and is at the same time transformed in P', while the rule on the right means if $P \xrightarrow{\alpha} P'$ is true and $other..conditions$ are satisfied, then $Q \xrightarrow{\beta} Q'$ is true. This kind of operational semantics has become a standard approach for defining process algebra semantics.

However, the most important semantic approach for process algebra is bisimulation (BS for short) semantics, including equivalence and congruence semantics based on bisimulation techniques. This approach follows the idea of behavioral equivalence. Two processes are considered undistinguishable if they can simulate each other when performing any actions. We would like to call this approach a comparative semantics approach. Usually, a bisimulation is also an equivalence relation, but not necessary a congruence relation, which requires two processes bisimulate in each context if they ever bisimulate. According to the different requirements of undistinguishability, different kinds of bisimulation have been developed, such as strong BS, weak BS, barbed BS, ground BS and open BS. A detailed discussion on these different BS can be found in [46], [42] or [41]. Lin has developed a theory and also model checker for symbolic BS[36]. Ying performd a thorough study on inexact BS, where he introduced a topological theory of processes in 2001[50] and investigated approximate BS, limit BS and near BS, and proposed BS index in 2002[51], with regard to processes in topological spaces. Recently, Ying even studied process algebra with noisy channels and introduced noisy $\pi$ calculus[52], where he defined approximate and stratified BS based on a probabilistic transitional semantics of $\pi$ calculus.

When Tony Hoare published the first version of his great work 'communicating sequential processes', no model or semantics is given[30]. Afterwards, in 1980, a model based on trace semantics was given for an enhanced version of CSP[31], the Theoretical CSP (TCSP). In this model, apart from functional improvements, Hoare used communication traces to characterize a denotational semantics of TCSP. It works with prefix-closed sets of traces and is thus only suitable for describing simple safety properties, but not suitable for complicated ones such as deadlocks. The detection of this disadvantage led to the introduction of failure semantics by Brookes, Hoare, and Roscoe[10]. In the new approach, each communication trace $\alpha$ is paired with a set X of events called a refusal set. The pair ($\alpha$ , X) is called a failure. It means that the process can first perform $\alpha$ and then refuse to perform any action in X, thus the name failure semantics. Later Brookes and Roscoe improved the failure model further to include divergence traces and proposed the failure/divergence model[13]. Recently, Brookes introduced the action trace semantics to unify the semantic description of shared memory parallelism, asynchronous and synchronous communication[14].

There has been also study on axiomatic semantics for CSP and CSP like languages. The first work in this direction was done by Apt in 1980[3]. Apt defined his axiom rules on the sets of communication records and obtained both partial and total correctness results. Apt's original work has some disadvantage that the pre-resp.-post conditions of an axiom contain only local variables of the reasoning process itself. Improvements were made in a subsequent paper in 1983[7]. On the other hand C. Zhou and Tony Hoare studied also partial correctness of communicating processes and protocols in 1981[53], followed by a compangion work of its scond author on total correctness of communicating processes[32]. A recent work on complete axiomatic semantics of CSP stable failures model was given by Yoshinao Isobe in 2006[33].

In summary, we see that people have proposed structural operational semantics, denotational semantics (in form of trace semantics), axiomatic semantics and comparative semantics (in form of bisimulation semantics) for different kinds of process algebra. Another line of research on formal semantics of process algebra is the algebraic approach. In next sub-section we will review the research work on algebraic specification techniques and their application on semantic specification of processes.

### 1.2 Heterogeneous algebra and abstract data types

The motivation of this paper is to introduce another kind of semantics for process algebra—the algebraic semantics based on deep isomorphism of seed algebras. In the search for a formal semantics that is more delicate than the denotational semantics, people have proposed a special kind of universal algebra, the heterogeneous algebra, as a tool for describing formal semantics of programs, usually called algebraic semantics. The syntactic framework of this semantics is signature, which presents an abstract syntax of the program. The essence of a signature is a set of algebraic rules charactering relations between data types called sorts. There are basic sorts like integer, real, Boolean etc. Examples of more complicated ones are queue, stack, array etc. Even a program can be considered as a sort. Context sensitivity can be introduced with a set of axioms, which is basically an equational system. A signature together with an axiom system is called an abstract data type (ADT). Note that the idea of ADT was born in early seventies and has gone a long way of development. ADT can be implemented with different techniques. What we mean here is a particular technique of ADT implementation, which is the combination of ADT idea with algebraic specification techniques and many sorted logics. We still call it ADT. Examples of early important contributions to ADT research are [54], [27], [24], [15], for more reference see [47]. The advantage of using ADT as algebraic semantics is that it provides not only a single model of semantics for a given program, but a set of different models, depending on different choice of axiom systems. Each such model is called a $\Sigma$ algebra, whose elements belong to different carriers. There is a one-one correspondence between the set of carriers and the set of sorts. Thus the $\Sigma$ algebras are heterogeneous. These models may form a mathematical structure, eg. a lattice. They even form a category. Different models represent different semantics. This is the advantage of algebraic semantics towards other kinds of semantics (operational, denotational, axiomatic). The mostly preferred model is the initial model (initial algebra), which is defined as the the quotient algebra $T(\Sigma)/E$ , where $T(\Sigma)$ is the ground term algebra of signature $\Sigma$ and E is the axiom system of ADT. Thus the initial algebra is the 'finest' model among all models and represents the initial semantics[28]. The condition for its existence is very weak. In case that the axiom system is equational, the initial algebra always exists. The initial algebra, if it exists at all, is unique up to isomorphism. For each model M of the category, there is a homomorphism h (morphism if we use the language of category) from the initial algebra to M, which is unique up to isomorphism.

Just as said by Wirsing: "Data stractures, algorithms and programming languages can be described in a uniform implementation independent way by axiomatic abstract data types"[48]. The idea of considering programming languages as ADT first appeared in 1980, where Broy & Wirsing[16] and Cleaveland[21] proposed this idea separately almost at the same time. To describe semantics of programs, which do not necessary terminate, Broy and Wirsing introduced the concept of partial heterogeneous abstract data types together with the conepts of weak and coweak homomorphisms between partial $\Sigma$ algebras[19]. In another paper[18], Broy considered the algebraic semantics of a small language and proved that its models do not form a complete lattice. That means that it neither has an initial model nor a terminal one. Instead, it possesses a coweak initial model and a weak terminal model. Ruqian Lu did a step further by considering algebraic semantics of progams with dynamic structures containing goto statements and nesting program blocks[38]. By introducing the concepts of program pointer and program running history in the ADT, Lu succeeded in providing the algebraic semantics for programs with dynamic structures.

Later, ADT as a mathematical tool of algebraic specification has been used in other areas, such as software engineering[48], data bases[22], Web services[23], operating systems[35], electronic payment[26], etc.

However, the application of ADT techniques to define process algebra semantics was not always successful. In a joint paper Astesiano, Giovini and Reggio proposed an observational semantics and logic[5]. They tried to give a unified and abstract semantic framework for concurrent systems and provided an abstract version of the Henessy Milner logic. Their basic motivation 'the integration of process specifications into the general schema of algebraic specifications of abstract data types' was very close to the motivation of our paper. However, the authors found that 'it is rather well known that the classical notions of semantics for algebraic specifications turn out to be not adequate for expressing sensible semantics for processes'. In another paper, Astesiano and Reggio, while making a survey on algebraic specification techniques for concurrency structures, repeated this statement and pointed out that for example the semantics given in [8] can not prove the equivalence of nil and nil + nil. They emphasized further that 'also the common form of observational semantics for algebraic specification ……allows only to express particular bisimulation semantics'[8]. All these considerations motivated them to develop a highly abstract approach of process semantics based on algebraic specification, which is formalism independent and only towards abstract concurrency models. In yet another paper of Astesiano, Mascari, Reggio and Wirsing[6], the authors proposed to use ASL, an algebraic specification technique with parametrized data types, to describe concurrency semantics. Also in this paper no concrete discussion on bisimulation techinqieus was given. Because of this reason they did not give detailed technical treatment of various concrete bisimulations, nor did they consider the relationship between these bisimulations. This is just what we want to do in this paper.

Later, a comprehensive survey on different kinds of algebraic specification techniques on concurrent systems was given by Astesiano, Broy and Reggio[2]. This paper can be considered as an extended and updated version of [8]. They discussed these techniques in a general setting, including models such as labeled transition systems; formalisms such as process calculi and algebras like CCS and CSP; application tools like Lotos and PSF. Although the paper is long and comprehensive, technical details such as various bisimulation techniques and their semantic description in algebraic style are not discussed.

### 1.3 Main contribution of this paper

The main goal of this paper is to establish an algebraic semantics of process algebra in form of ADT and to build a bridge between this algebraic semantics and the traditional operational and bisimulation semantics. The main result of this effort is a set of equivalence theorems stating that two processes (in some sense) bisimulate each other if and only if their corresponding seed algebras (a specific form of $\Sigma$ algebra) are deep isomorphic. Note that although the study on programming languages as ADT have been discussed since long ago, but such an equivalence between algebra isomorphism and bisimulation of process algebras has (to our knowledge) never been discussed.

There are three major problems that play a key role in obtaining the above result. The first one is about the form of the $\Sigma$ algebra, which corresponds to the behavior of a process such that isomorphism of two $\Sigma$ algebras is equivalent to bisimulation of two processes. Certainly this $\Sigma$ algebra is not the initial algebra of the ADT, since different processes are built on the same ADT. The initial algebra of an ADT is unique up to isomorphism. It would be meaningless to let all processes correspond to the same initial algebra. We propose instead the concept of seed algebra, where the seed can be any defined element of the initial algebra. A seed algebra is a subalgebra of the ADT's initial algebra such that it is 'finitely generated' from the seed. (Note that the initial algebra is 'finitely generated' from all elements of the carrier sets) For the purpose of describing the behavior of processes, we let the seed to be the process itself or the process together with its initial state. The signature of the ADT is designed in the way such that only operations representing process transitions (and their associate operations, such as choose the left or choose the right) are applicable to the seed algebra elements (Later it will be generalized to process behavior other than transition, e.g. barb check in a barbed bisimulation). It is then easy to see that the elements of a seed algebra form just the set of reachable processe states of the original process, i.e. all processe states reachable from the seed by transitions. It is in this sense that we call a seed algebra, where the seed is a process or the initial state of the process, the algebra of process behavior.

The second major problem is how to define the isomorphism of two seed algebras. This problem is interesting since it is made difficult by the non-deterministic character of process behavior. Consider the processes P = ***a*** | ***b***, Q = ***a.b*** + ***b.a*** + ***a*** | ***b***. Obviously P and Q bisimulate each other. When P performs action ***a***, Q may choose either the first or the third summand to perform the same action. Similarly, if P performs action b, Q may choose either the second or the third summand to perform the same action. This means we have at least two difficulties. The first difficulty is that the correspondence of action selection is not one-to-one. This implies that the correspondence between elements of the two seed algebras (with P or Q as seed) is not one-to-one, which violates the definition of isomorphism. The second difficulty is how to represent the implicit decision 'select left summand' in a sum process or 'select right component' in a composition process with our seed algebra. Here we are entering a dilemma. If we do not specify these implicit decisions in the representation of seed algebra elements, then ambiguity appears. But if we do specify them, then bisimilar processes would correspond to non-isomorphic seed algebras, since the seed algebra of Q includes the information 'select left summand', but the seed algebra of P does not. These difficulties confirmed the statement made by Astesiano[5, 8] on the inappropriateness of traditional algebraic specification techniques for specifying formal semantics of process algebra from a particular aspect, the bisimulation aspect. The solution we propose in this paper is to differentiate between hidden and non-hidden operations, and define the rough seed isomorphism only as isomorphism of non-hidden operations. All implicit decisions made to resolve non-determinism in process algebra will be programmed as hidden operations, which will be neglected in rough homomorphism or isomorphism. On the other hand, each action (in particular, observable action) of a process will be represented by a non-hidden operation (which represents the action itself), possibly preceded by a sequence of the algebra's hidden operations (which represent the non-deterministic selection procedure). For the above example, let's use 'left' and 'right' to denote the non-deterministic choice, and 'act' to denote the performance of an action, then the algebraic expression act (a, left (p)) may denote the transition of first selecting the left component of P and then performing the action 'a' in the prefix of this left component, where p = compo (a, b) is the algebraic expression of P. Now we want Q to simulate P. But there are two possibilities for Q to perform the action 'a'. The corresponding algebraic expressions can be act (a, left (q)) or act (a, left (right (right (q)))), where q is written in the nested form q = sum (a.b, sum (b.a, compo (a, b))). This shows that for performing action 'a' the correspondence {act (a, left (p))}

←→ { act (a, left (q)), act (a, left (right (right (q))))} is a 1-to-2 correspondence. We are not happy with it and want to have a 1-to-1 correspondence. How can we achieve that? We let 'left' and 'right' be hidden functions and 'act' be a non-hidden function. Roughly, we define the isomorphism as one of non-hidden functions only and ignore all hidden functions. We summarize the hidden and non-hidden functions of both sides and rewrite the above correspondence as $\overline{act}$([a], [p]) ←→ $\overline{act}$([a], [q]), where $\overline{act}$([a], [p]) = {act (a, left (p))} and $\overline{act}$([a], [q]) = { act (a, left (q)), act (a, left (right (right (q)))) }. Now we see that the correspondence becomes 1-to-1, where [p] is called the hidden extention of p and $\overline{act}$([a], [p]) is the non-hidden closure of [a] and [p], etc. This is a 1-1 correspondence of non-hidden closures of both seed algebras. Two seed algebras are called rough seed isomorphic if there is a rough seed isomorphism between their non-hidden closures. This shows how we use hidden technique to redefine the isomorphism between $\Sigma$ algebras. Another use of hidden techniques in this paper is to resolve the difficulty of unbalanced actions between two weakly bisimulating processes. We know that for weak bisimulation it is possible that one process performs a $\tau$ action, while the other process performs either nothing, or a $\tau$ action, or also more than one $\tau$ action. In order to keep 1-to-1 correspondence between unbalanced actions, we need the help of hidden functions to represent different process behavior 'implicitely', just like we use hidden functions to represent non-determinism 'implicitely'. For details see section 5. Our paper proves that two processes are (strongly) trace equivalent (having the same set of traces) if and only if their seed algebras are rough seed isomorphic. This condition is not enough for bisimulation. We prove further that two processes are (in some sense) bisimular if and only if their seed algebras are deep isomorphic. We will see that different bisimilarity theorems need different axiom systems.

The third major problem is how to represent and prove the relationship between different bisimulations. We present an ADT Process-Trans in section 3-4 to represent strong early bisimulation as deep isomorphism of seed algebras of this ADT. However, it would be cumbersome to write all different kinds of bisimulations in this way and to prove the correspondence of bisimulation and deep isomorphism one by one separately. Rather we make use of modularity of abstract data types and build all bisimulations as a hierarchical tree of ADT. This approach has the advantages of 1. avoiding writing many parts of signatures and axioms redundantly; 2. using inheritance of abstract data types to represent ordering of bisimulation power; 3. combining the proof of correspondence: bisimulation ←→ deep isomorphism with the proof of bisimulation power ordering in one step.

The remaining part of this paper is organized as follows. In section 2 we introduce the ADT as a tool of describing algebraic semantics for process algebra. In order to avoid involving too many details at once, we first define a very simple process algebra and discuss the property of its ADT. Then we gradually enrich the process algebra with more and more features. Each time when introducing new grammar features of process algebra we also introduce and prove new properties of the ADT. In this section we also introduce the concept of seed algebra, which is the fundamental concept in this paper. In section 3 we introduce rough seed isomorphism of seed algebras. We will prove that rough seed isomorphism is enough for characterizing (strong) trace equivalence, but not enough for characterizing (strong) bisimulation. In section 4 we will introduce deep seed isomorphism and prove that this kind of isomorphism is characteristic for process bisimulation. We provide an ADT with an axiom system, whose deep isomorphism characterizes early strong bisimulation of processes based on a process grammar with input, output, communication and recursion. In section 5 we extend the results to prove the equivalence between deep isomorphism and other types of bisimulation. We use a hierarchical data type to describe the relationship between different types of bisimulation and to prove this relationship, which is a directed acyclic graph containing ten different bisimulation types. Each node denotes a bisimulation type and each directed arc from A to B means that type A bisimulation is weaker than type B bisimulation (or, type B bisimulation is a sub-relation of type A bisimulation). Weak barbed bisimulation is at the weakest end and strong open bisimulation is at the strongest end of this graph. This shows that the relationship between different bisimulation types can be completely characterized with different models of abstract data type semantics. Finally, in section 6, we conclude the whole work with some remarks on the future work.

## $2 Hierarchical ADT and seed algebra

For giving algebraic semantics to process algebras, we have to introduce several new definitions to enrich the traditional idea of abstract ADT as programming language semantics. These new definitions will be introduced in the sequel in a stepwise way.

**Example 2.1**: First let us consider a very simple process algebra. Its processes are defined by the following rule:

P ::= nil | a. P | b. P (2.1)

This process algebra has a simple operational semantics:

$$\alpha.P \xrightarrow{\alpha} P, \quad \text{where } \alpha \in \{a, b\} \qquad (2.2)$$

In order to use $\Sigma$ algebra to describe the structure and behavior of this process algebra, consider that we only need two sorts **name** (for action symbols) and **proc** (for processes). The action symbols a and b, together with the zero process nil, form the set of constant operations (with no parameters). The other two operations are pre (used for structuring the processes) and act (used for representing the transition). Together with the axiom of the ADT, it defines the operational semantics given above. In summary, we have the following ADT:

**Example 2.2**:

**Type** Simple Process Algebra =

{**sort name, proc**

**op**

a: → **name**

b: → **name**

nil: → **proc**

pre: **name** × **proc** → **proc**

act : **name** × **proc** → **proc**

**axiom**

act (x, pre (x, t)) = t }

where pre (x, p) denotes a process x.p consisting of a prefix x and a process p, act (x, pre (x, p)) denotes a transition that removes the prefix x from pre (x, p), the only axiom says nothing else than the transition rule (2.2) above.

The heterogeneous algebras, which can be interpreted as $\Sigma$ algebras of this ADT, are many. For the sake of simplicity, we only consider its ground term algebra. The ground term algebra GT (Simple Process Algebra) = (A, F), where A is the set of its elements and F the set of its functions, consists of the following:

$A = \{A_{name} = \{a, b\},$

$A_{proc}$ = {nil, pre (a, nil), pre (b, nil), pre (a, pre (a, nil)), pre (a, pre (b, nil)), ….

act (a, nil), act (b, nil), act (a, act (a, nil)), …..

act (a, pre (a, nil)), act (a, pre (b, nil)), ….

pre (a, act (a, nil)), pre (a, act (b, nil)), ….}}

F = {a, b, nil, pre, act},

where $A_{name}$ is the carrier set of elements of **sort name** (i.e. all elements of **sort name)** and $A_{proc}$ the carrier set of elements of **sort proc,** which has infinitely many elements.

Since our motivation is to make use of the idea of $\Sigma$ algebra to give an algebraic semantics for process algebra, we will investigate this example a little bit in detail to check what is missing in conventional $\Sigma$ algebra that is needed to describe the semantics of process algebra.

Comparing the traditional $\Sigma$ algebra theory (and the above example), we have found at least two problems.

The first problem is about the suitability of the $\Sigma$ algebra as a representation of the process set. The elements of this ground term algebra include symbolic representations of all processes that can be generated by the process grammar (2.1) above. They include also representations of all processes that are produced by transitions of other processes (via the act operation). But this is not yet all. They do far much more. The ground term algebra also includes elements that represent impossible transitions like act (a, nil) or act (a, pre (b, nil)), which are not representations of any transitions. In other words, they should be invalid elements of the $\Sigma$ algebra according to some axiom system.

But even if we consider only those elements of the ground term algebra that are legal representations of processes, they are still not appropriate for describing process algebra. Just as we said above, this ground term algebra covers both the syntactical and semantic aspects of process algebra. Syntactically, it shows how the processes are constructed from their basic elements a, b, nil and their basic construction tool: the pre operation. Semantically, it uses the act operation (together with the axiom) to reflect the transition rule of a process. This way of mixing syntax with semantics together brings in the problem that syntactical and semantic constructs can be applied to algebra elements in an alternating way. The result can be then semantically meaningless. Consider the element

pre (a, act (b, pre (b, nil)))   (2.3)

By equivalence, this is a legal representation of a process, since

pre (a, act (b, pre (b, nil))) = pre (a, nil)

But by intuition, only the right hand side of the above equation is meaningful, while the left hand side is meaningless since it would mean that one first constructs a process pre (b, nil), then lets it perform a transition (do the action b) to become nil, then uses the result to construct a new process pre (a, nil). This is impossible because it mixes the program construction stage with the program running stage.

In order to solve these two and further problems, we introduce four mechanisms. The first one is to design a hierarchy of layers of ADT. We divide each ADT in a hierarchy of layers. The first layer is the syntactic layer, where the signature is constructed to match the recursive definition of process grammar. This is also the very basic layer of the hierarchy. All other layers form a tree with this basic layer as the root. Any node of the tree, except the basic layer, represents a particular bisimulation. Each node (layer) inherits sorts, operations and axioms from its father node (layer), possibly enriched (using notation +) with additional sorts, operations and axioms. This hierarchy helps us to explore the relationship between different bisimulations. The second new mechanism is to introduce a valve of inheritance between two consecutive layers of hierarchy of ADT. This is the **private** attribute for operations. Operations of the lower layer, which are not intended to be used by the higher layer, will be marked as **private.** That means they are inaccessible to all higher layers. In this way the (static) construction and (dynamic) transition of processes will not be mixed together. Another means for implementing a valve of inheritance is to use the minus notation (-) to refuse the inheritance of some features. While the **private** attribute is a property of the lower layer, the minus notation (-) is a property of the higher level, which is used when some, but not all, higher levels want to inherit some features from the same lower level. The third one is to introduce another attribute for operations: the **hidden** operations. We have given the reasons of introducing hidden operations above and will discuss its details in section 3 (rough seed isomorphism) and in section 5 (unbalanced active/passive actions of bisimulation). The fourth one is to divide the $\Sigma$ algebra in sub-algebras such that each sub-algebra describes the behavior of some particular process. In this way we can differentiate and compare behaviors of different processes. As it will be seen later, we call this kind of subalgebra seed algebra.

**Definition 2.1**: A general syntax of ADT is as follows:
<ADT> ::= <Basic Layer> | <Extended Layer>
<Basic Layer> ::= **Type** <Basic Layer Name> = {<B-Layer Body>}**End of** <Basic Layer Name>
<Extended Layer> ::= **Type** <Extended Layer Name> =
{<E-Layer Body>}**End of** <Extended Layer Name>
<B-Layer Body> ::= <Signature> <Axiom Part>
<Signature> ::= <Sort Part> <Operation Part>
<Sort Part> ::= {**sort** {<Sort Name>}$_1^n$ }$_0^1$
<Operation Part> ::= {**op** {<Oparation>}$_1^n$ }$_0^1$ {**private op** {<Operation>}$_1^n$}$_0^1$
{**hop** {<Operation>}$_1^n$ }$_0^1$ {**private hop** {<Operation>}$_1^n$}$_0^1$
<Axiom Part> ::= {**axiom** {<Axiom>}$_1^n$ }$_0^1$
<E-Layer Body> ::= **Type** <Underlying Layer Name> <E-Signature> <E-Axiom>
<Underlying Layer Name> ::= <Basic Layer Name> | <Extended Layer Name>
<E-Signature> ::= <E-Sort> <E-Operation>
<E-Sort> ::= {+/- **sort** {<Sort Name>}$_1^n$ }$_0^1$
<E-Operation> ::= {+/- **op** {<Oparation>}$_1^n$ }$_0^1$ {+/- **private op** {<Operation>}$_1^n$}$_0^1$
{+/- **hop** {<Operation>}$_1^n$ }$_0^1$ {+/- **private hop** {<Operation>}$_1^n$}$_0^1$
<E-Axiom> ::= {+/- **axiom** {<Axiom>}$_1^n$ }$_0^1$
where for each X, {+/- X}$_0^1$ means {+ X}$_0^1${- X}$_0^1$. The syntax of <Operation> and <Axiom> is the same as it is given in literature, where operations headed by **hop** are hidden operations while those headed by **op** are non-hidden operations. Correspondingly, the interpretation of a hidden (non-hidden) operation in some $\Sigma$ algebra of this data type is called a hidden (non-hidden) function. As result, we get a hierarchy of ADT. Each extended layer inherits everything from the underlying layer in a default way, unless it refuses to inherit something (by using the minus symbol -), and at the same time introduces some new things (by using the addition symbol +)

If some operation f of the underlying ADT is declared as **private** then f cannot be applied in the extended ADT to any term t, which is not a term of the underlying ADT. (More precisely, f cannot use elements that only belong to extended ADT but not belong to underlying ADT, to construct new elements). For a more practical rule of applying the **private** property to term construction in ADT see definition 2.2.

**Example 2.3**: A small ADT:
**Type** First =
{**sort** s1
**op**
a: → s1
**private op**
exp: s1 → s1
}
**End of Type** First

**Type** Second =
{**Type** First
+ **sort** s2
+ **op**
c: → s2
d: → s2
add: s2 × s2 → s1
}
**End of Type** Second

According to the meaning of a private operation, exp (add (c, d)) would be an illegal element, since it applies a private operation exp of Type First to the elements c and d of Type Second, which do not belong to Type First.

We introduce recall some well known definitions for the convenience of the reader.
**Definition** 2.2: A $\Sigma$ algebra of the signature is a pair (A, F), where A is an indexed set of carrier sets: A = { $A_{s_i}$ | $s_i$ is a sort of the signature}, where all elements (called ground terms) of $A_{s_i}$ have the sort $\mathbf{s_i}$. F is the set of functions: F = {f | f is the interpretation of some operation o of the signature}. For each operation:

o: $s_1 \times s_2 \times ... \times s_n \to s_{n+1}$,

each function f interpreting o, and each n-tuple ($a_1$, …, $a_n$), where for all i, $a_i \in A_{s_i}$, we have $f(a_1,...,a_n) \in A_{s_{n+1}}$, where o is either defined as a non-private operation in some underlying

ADT, or is defined in the current signature as an enrichment to the underlying ADT.

**Definition** 2.3: If in addition for all i, the set of all constant functions (i.e. having functionality zero) in $A_{s_i}$ is just those constant operations given in the signature, then the algebra is called the ground term algebra of the ADT. We denote it with GT (D) if D is the name of the ADT. This is also true if D has a non empty axiom part.

**Definition**:2.4: If in addition each $A_{s_i}$ is allowed to contain a subset of variable terms of sort $\mathbf{s_i}$, which are either variable names or function elements $f(t_1,...,t_n)$ with parameters $t_1,...,t_n$, which are at least partly variable terms, then the algebra constructed in the same way as definition 2.3 is called term algebra and denoted with T (D).

**Definition** 2.5: An axiom has the form e1 = e2, where e1 and e2 are both (ground or variable) terms of the term algebra. The axioms form an equational reasoning system.

**Definition** 2.6: If ADT D has an axiom part AX, then the quotient algebra GT (D) / AX is called the initial algebra IN (D) of D. It is also called a model of the ADT.

**Definition** 2.7: Given ADT D and its two $\Sigma$ algebras A ($\Sigma$) = (A, F) and B ($\Sigma$) = (B, G), where A = { $A_{s_i}$ } and B = { $B_{s_i}$ } are the carrier sets, F = {$f_i$} and G = {$g_i$} are the function sets, B ($\Sigma$) is called a subalgebra of A ($\Sigma$), if for each operation symbol Op, and each $g_i \in$ G interpreting Op, there is a $f_i \in$ F also interpreting Op, such that g = f $|_B$。

**Definition** 2.8: Given ADT D and its two $\Sigma$ algebras A ($\Sigma$) = (A, F) and B ($\Sigma$) = (B, G) as above. A mapping $\varphi$: A ($\Sigma$) → B ($\Sigma$) is called a $\Sigma$ homomorphism, or simply homomorphism, if the following requirements are satisfied:

1. For all $a \in A_{s_i}$, $\varphi(a) \in B_{s_i}$,

2. For any operation f of the signature, if $f_A(a_1,...,a_n) = a_{k+1}$, where $f_A$ is the interpretation of f in F, $\forall i$, $a_i \in A_{s_i}$, there is $f_B(b_1,...,b_n) = b_{k+1}$, where $f_B$ is the interpretation of f in G, $b_i \in B_{s_i}$ such that $\varphi(f_A(a_1,...,a_n)) = f_B(\varphi(a_1),...,\varphi(a_n))$.

$\varphi$ is called an isomorphism if it can be reversed. (one-to-one or bijection.)

Now we will improve the ADT given in example 2.2 and provide a new one based on above principle of hierarchical ADT:

**Example** 2.4:

**Type** Simple Construct =

{**sort name, proc**

**op**

a: → **name**

b: → **name**

nil: → **proc**

**private op**

pre: **name** × **proc** → **proc }**

**Type** Simple Process Algebra =

{**Type** Simple Construct +

**op**

act : **name** × **proc** → **proc**

**axiom**

act (x, pre (x, t)) = t

}

In example 2.4, we have a hierarchy of two data types. The lower level data type, Simple Construct or $D_1$, is responsible for constructing all processes and plays the same role as the process grammar given at the beginning of this section. The higher level data type, Simple Process Algebra or $D_2$, mimics the transition procedure of processes generated by $D_1$. The operation symbol 'pre' is private and does not belong to F ($D_2$). That means it is not allowed to apply the operation 'pre' to those elements, which only belong to A ($D_2$) but do not belong to A ($D_1$). Now we see that A ($D_2$) will no more contain ill-structured algebra elements like pre (a, act (a, pre (a, nil))). That means the data type will no more mix the process construction procedure with the process transition procedure.

There is still another problem when defining $\Sigma$ algebras of our ADT. This is the problem of how to describe the context sensitivity. We need means for preventing illegal algebra elements. We have already solved the problem of preventing undesirable elements like pre (a, act (b, pre (b, nil))) in the above example by introducing hierarchical ADT and private (uninheritable) features of ADT to avoid unwanted mixture of features from different layers. However, there are other risks of introducing undesirable elements such as act (a, pre (b, nil)) which would correspond to impossible transition like $b.nil \xrightarrow{a} ?$. This means that we have to take the problem of defining 'undefinedness' in consideration. In the literature, $\Sigma$ algebra containing undefined elements is often called partial algebra. One approach of dealing with partial algebra is to consider it as a special kind of total algebra and to consider undefined elements as legal elements of the algebra. Then one can define different kinds of algebra homomorphisms called weak homomorphism and coweak homomorphism[16]. In this paper, we define the undefinedness implicitely with context conditions. Our approach has the advantage that we do not need to write down a long list of undefined elements. Another advantage of our approach is that we do not have any need to discuss partially defined models (models that include some undefined elements but exclude the others). All seed algebras contain only defined elements.

Except some isolate conditional equations, the axiom system is a set of equations with variables. It is written for the term algebra. However, what we are going to do is to introduce context sensitivity in ground term algebra. Therefore, when we say that some element is undefined, we are talking about elements of the ground term algebra.

The following definition is about undefinedness.

**Definition** 2.9

We assume that D is an ADT with sorts, operations and axioms. We assume also

$$f: \mathbf{s_1} \times \mathbf{s_2} \times \ldots \times \mathbf{s_i} \times \ldots \times \mathbf{s_n} \rightarrow \mathbf{s_{n+1}}$$

where n > 0, is any operation of D. Consider an axiom system where all terms belong to the term algebra. Then we have the following rules of definedness:

1. All terms (and subterms) appearing in an axiom system are source terms;
2. Term T' is equivalent to T if T' can be obtained from T by equational reasoning with the axiom system;
3. If term T is a source term, T' is a subterm of T, then T remains to be a source term if its subterm T' is replaced by an equivalent subterm T'';
4. Term f ($t_1$, $t_2$, …, $t_n$) is undefined if it is not an instantiation of at least one source term;
5. Term f ($t_1$, $t_2$, …, $t_n$) is undefined if at least one of the $t_i$ is undefined;
6. If term T is undefined, T' can be obtained from T by equational reasoning with the axiom system, then T' is undefined, too;
7. All other terms are defined.

Note that an instatiation of a source term is not necessary defined. Moreover, even if all $t_1$, $t_2$, …, $t_n$ are defined, f ($t_1$, $t_2$, …, $t_n$) is not necessarily defined. Our definition of undefinedness is context sensitivity oriented.

**Example2.5**: Consider the axiom system given in Appendix 1.

The following elements are undefined:

tinput (tconf (o-proc (c, a, nil) , tp, ()), c, b)

toutput (tconf (i-proc (c, a, nil) , tp, ()), c, a)

tinput (tconf (nil, tp, ()), c, a)

tleft (toutput (tconf (o-proc (c, a, nil), tp, ()), c, a))

We only show the undefinedness of the first element

tinput (tconf (o-proc (c, a, nil), tp, ()), c, b) is undefined, because its equivalent term

tconf (input (o-proc (c, a, nil), c, b), tp, run ((), ot (c, b))) is undefined, because its subterm

input (o-proc (c, a, nil), c, b) is undefined, because

it is not an instatiation of any source term

**Definition 2.10:** Assume that some ADT D and its grund term algebra GT (D) are given. Further assume ***a*** is a defined element of GT (D), where ***a*** is either a constant function or ***a*** = g ($\boldsymbol{b_1,\ldots,b_n}$), where each $\boldsymbol{b_i}$ is an element of GT (D) and g is a non-hidden function. The loose seed algebra LA (D, ***a***) is defined as follows:

1. $\boldsymbol{a} \in$ LA (D, ***a***);
2. If for some *i*, $b_i \in$ LA (D, ***a***) and $f(b_1,\ldots,b_i,\ldots b_n)$ is defined, then $f(b_1,\ldots,b_i,\ldots b_n) \in$ LA (D, ***a***), where

$$f: \mathbf{s_1} \times \mathbf{s_2} \times \ldots \times \mathbf{s_i} \times \ldots \times \mathbf{s_n} \rightarrow \mathbf{s_{n+1}}$$

is an operation of D, all $b_1,\ldots,b_i,\ldots b_n \in$ GT (D), each $b_j$ is of sort $\mathbf{s_j}$.

Then the seed algebra A (D, ***a***) is defined as the quotient algebra LA (D, ***a***) / {all axioms}. ***a*** is called the seed of A (D, ***a***).

**Proposition 2.1:** Let GT (D) be the ground term algebra of D, ***a*** is an element satisfying the requirement of definition 2.10, then

1. LA (D, ***a***) is a (partial) subalgebra of GT (D);
2. A (D, ***a***) is a (partial) sub-algebra of IN (D).

Proof: 1. We prove two facts. First, the set of all elements of LA (D, ***a***) is a subset of the set of all elements of GT (D), since the seed belongs to GT (D) and all operations applied to elements of the loose seed algebra map these elements to elements of GT (D). Second, the elements of the loose seed algebra really form a subalgebra of GT (D), since for any operation of GT (D):

$$f: \quad \mathbf{s_1} \times \ldots \times \mathbf{s_i} \times \ldots \times \mathbf{s_n} \rightarrow \mathbf{s_{n+1}}$$

and elements $(a_1 : \mathbf{s_1}, \ldots, a_n : \mathbf{s_n}) \subseteq$ LA (D, ***a***),

a) either f $(a_1, \ldots, a_n)$ is undefined,

b) or f $(a_1, \ldots, a_n)$ is defined. In this case f $(a_1, \ldots, a_n) \in$ LA (D, ***a***), since all $a_i$ contain the seed as their subterm, which implies that f $(a_1, \ldots, a_n)$ contains the seed as its subterm, which in turn implies that f $(a_1, \ldots, a_n)$ belongs to LA (D, ***a***) according to the definition of loose seed algebra.

2. Note that the initial $\Sigma$ algebra IN (D) of D is the quotient algebra GT(D) / {all axioms}. The second conclusion of this proposition is true according to a general reasoning that if algebra A is a subalgebra of algebra B then the quotient algebra A/ AX is a subalgebra of the quotient algebra B / AX, where AX is an axiom system.

□

Note that A (D, ***a***) may also be a proper sub-algebra of IN (D). In fact, generally speaking, in most of the cases, all seed algebras are proper subalgebras.

**Example 2.6**: Consider the ADT:

D =
{**sort water, fire**
**op**
drop: → **water**
spark: → **fire**
drink: **water** → **water**
light: **fire** → **fire**
}

In the ADT D the axiom set is empty. The ground term algebra GT (D) contains infinitely many different seed algebras: A (D, drop) = {drop, drink (drop), drink (drink (drop)), …}; A (D, drink (drop)) = {drink (drop), drink (drink (drop)), …};…..;A (D, spark) = {spark, light (spark), light (light (spark)), …}; A (D, light (spark)) = { light (spark), light (light (spark)),…};…… In this example, there is no seed algebra which equals to the ground term algebra ( where the axiom set is empty)

Another fact is that for fixed D and ***a***, the seed algebra A (D, ***a***) is uniquely defined. This conclusion can be proved with structural induction.

The next step is to discuss the relation between different seed algebras. We may want to introduce homomorphisms and isomorphisms between two seed algebras. We call them seed homomorphisms (isomorphisms). However, we will see that this concept is not very fruitful as it will be seen from the discussion below.

**Definition 2.11:** Given two seed algebras SA1 = A (D, e1) and SA2 = A (D, e2). A seed homomorphism (isomorphism) $\varphi$ from SA1 to SA2 is a usual $\Sigma$ algebra homomorphism (isomorphism) with the additional rule $\varphi$ (e1) = e2.

It is worthwhile to note that the above definition implies that in any such homomorphism defined (undefined) elements should be mapped to defined (undefined) elements.

**Example** 2.7: Consider the hierarchical data type in example 2.4 and its ground term algebra.

A ($D_2$) = { $A_{name}$ = {***a***, ***b***},

$A_{proc}$ = {nil, pre (***a***, nil), pre (***b***, nil), pre (***a***, pre (***a***, nil)), pre (***a***, pre (***b***, nil)), …….,

act (a, nil), act (b, nil), act (a, pre (a, nil)), act (a, pre (b, nil)), act (b, pre (a, nil)), act (b, pre (b, nil)), …..}

}

F ($D_2$) = {***a***, ***b***, nil, pre, act}

where F ($D_2$) is the set of functions, $D_2$ = Simple Process Algebra.

It is easy to see that this ground term algebra contains many 'nonsense' terms. Only when we take the axiom act (x, pre (x, t)) = t of $D_2$ in consideration we will achieve two things. First, we will find out the 'nonsense' terms, which are those who are judged to be undefined by using definition 2.9 above. This step will remove the elements like act (a, nil), act (b, nil), act (a, pre (b, nil)), act (b, pre (a, nil)), etc. Second, we calculate the quotient algebra of GT ($D_2$) with respect to this axiom and obtain the initial algebra IN ($D_2$). This step will remove all elements starting with 'act', because any such element, when it is defined, is equivalent to some element without 'act'.

With this preview we go ahead with our discussion with seed algebras. Let the two seeds be:

e1 = pre (***a***, pre (***b***, nil), e2 = pre (***b***, pre (***a***, nil))

Then we will have two different seed algebras.

A ($D_2$, e1) = A ($D_2$, pre (***a***, pre (***b***, nil)) = {nil, pre (***b***, nil), pre (***a***, pre (***b***, nil))}

A ($D_2$, e2) = A ($D_2$, pre (***b***, pre (***a***, nil))) = {nil, pre (***a***, nil), pre (***b***, pre (***a***, nil))}

A simple isomorphism would be

$\varphi$ (***a***) = ***b***,

$\varphi$ (***b***) = ***a***,

$\varphi$ (nil) = nil,

$\varphi$ (pre(***b***, nil)) = pre ( $\varphi$ (***b***), $\varphi$ (nil)) = pre(***a***, nil),

$\varphi$ ( pre (***a***, pre (***b***, nil)) = pre ( $\varphi$ (***a***), pre ( $\varphi$ (***b***), $\varphi$ (nil))) = pre(***b***, pre (***a***, nil)).

**Example 2.7:** Given an ADT and two seeds:

1. Seed homomorphism between two seed algebras needs not to be unique;
2. Seed homomorphism between two seed algebras needs not to be an isomorphism;
3. Seed homomorphism between two seed algebras does not have to exist.

To see that these conclusions are true, we note the following:

1. For Example 2.7 given above, if there were three constant operations a, b and c, then we would have two possibilities:

$\varphi$ (***a***) = ***b***, $\varphi$ (***b***) = ***c***, $\varphi$ (***c***) = ***a***, or

$\varphi$ (***a***) = ***c***, $\varphi$ (***c***) = ***b***, $\varphi$ (***b***) = ***a***,

2. For Example 2.7 given above, if the axiom was act (b, pre (b, t)) = t, then we would have act (a, pre (a, t)) $\neq$ t. This means we only have a seed homomorphism from A ($D_2$, e1) to A ($D_2$, e2), but not vice versa.

3. Let e1 = pre (a, pre (b, nil)), e2 = nil. Then there is no homomorphism between A ($D_2$, e1) and A ($D_2$, e2). Otherwise, assume there was one such homomorphism $\varphi$, then it would be:

$\varphi$( pre (***a***, pre (***b***, nil)) = nil,

$\varphi$( act (a, pre (***a***, pre (***b***, nil)))) = act ($\varphi$(***a***), $\varphi$( pre (***a***, pre (***b***, nil)))) = act ($\varphi$(***a***), nil)

It is obvious that act ($\varphi$(***a***), nil) is undefined no matter whether $\varphi$(***a***) = ***a*** or $\varphi$(***a***)= ***b***. This means that the homomorphic mapping $\varphi$ does not exist.

□

From the discussion of this section we see that it is not appropriate to take the traditional concept of $\Sigma$ homomorphism as homomorphism of our seed algebras.

## $ 3 Hidden operation and rough seed isomorphisms

The motivation of introducing a new homomorphism (isomorphism) concept for seed algebras is to characterize bisimulation of processes. For this sake, it is necessary to find a solution for representing non-determinism and non-deterministic homomorphism. We will introduce algebraic mechanism with respect to ADT to describe non-determinism of process algebra transitions, namely non-determinism in sum operation (alternative choice) and non-determinism in parallel process composition. We will introduce hidden operations to make usually implicit non-deterministic choice explicit. We will introduce also rough seed homomorphism (rough seed isomorphism), which ignores all implicit non-deterministic choices in any homomorphic (isomorphic) mapping. In this way, we have found a technique to incorporate non-determinism in homomorphism (isomorphism) between two seed algebras.

Since our goal is to establish a correspondence between bisimulation of processes and isomorphism of seed algebras, we care only the action series performed by the processes, but not the non-deterministic selection procedure for determining which action to do next. For example, the sum process P + P is equivalent to the single process P, despite the fact, that for performing the same action series, the sum process must first select one of its summand processes. This selection procedure can be considered as a new kind of hidden actions (not the hidden action $\tau$, which is known in theory of process algebra) and should be differentiated from the other actions in the $\Sigma$ algebra. We call them hidden operations and mark the set of hidden operations with **hop**. It has all properties that a usual operation has with only one exception. That is, the hidden operation symbols are "irrelevant" in any rough seed homomorphism or isomorphism (to be explained later). Functions representing hidden operations are called hidden functions. Note that although there are hidden and non-hidden operations and functions, but the elements of seed algebra are not classified as hidden or non-hidden elements.

**Example 3.1**: Consider a new process grammar:

P ::= nil | x.P | P + P | (P | P)

x ::= a | b | c

The following hierarchical data type describes this process grammar:

**Type** Cons =

**{ sort name, proc**

**op**

```
   a, b, c :    → name
   nil :        → proc
   private op
   pre :      name × proc → proc
   sum :     proc × proc → proc
   compo :    proc × proc → proc
      }
Type Trans =
   {    Cons +
   op
   act:    name × proc → proc
   hop
   left :    proc → proc
   right:    proc → proc
   compol:    proc → proc
   compor:    proc → proc
   back:    proc → proc
   axiom
      left (sum (x, y)) = x
      right (sum (x, y)) = y
      left (compo (x, y)) = compol (x, y)
      right (compo (x, y)) = compor (x, y)
      left (compol (x, y)) = compol (left (x), y)
      right (compol (x, y)) = compol (right (x), y)
      left (compor (x, y)) = compor (x, left (y))
      right (compor (x, y)) = compor (x, right (y))
      act (z, pre (z, p)) = p
      act (z, compol (x, y)) = compol (act (z, x), y)
      act (compor (x, y)) = compor (x, act (z, y))}

      back (compol (x, y)) = compo (back (x), back (y))
      back (compor (x, y)) = compo (back (x), back (y))
      back (pre (z, p)) = pre (z, back (p))
      back (nil) = nil
      back (sum (x, y)) = sum (x, y)
      back (compo (x, y)) = compo (x, y)
   }
```

The base type Cons is used to construct the processes, where pre is used to construct a prefixed process, sum used to construct sum process, and compo used to construct parallel process composition. The hierarchical type Trans is used to describe transitions of processes, where the only non-hidden operation is act that is used to perform an action (i.e. delete a prefix). The hidden operations are used to remove non-determinism in the way that left operation selects the left component of a sum or compo structure, while right operation selects the right component. The case of sum operation is easier to deal with since after selecting a component the other component is discarded. The case of compo operation is more complicated. After selecting a component the other component still remains. Therefore we need something like a pointer to mark the place that we have selected. The compol operation marks the left component and compor operation marks the right component of a compo operation. Thus, each left (right) operation transforms a compo structure to a compol (compor) structure (see the axioms). If the marked component is itself a composite structure (sum or compo) the selection and marking procedure will be continued. We will get a nested structure of mixed compol and compor operations.

The axioms of the data type Trans form three groups. The first group includes all axioms involving the operation symbols ‘left’ and ‘right’.

**Example** 3.2    The transition

$$a.P + b.Q \mid c.R \xrightarrow{b} Q \mid c.R$$

where p, q, r represent the processes P, Q, R respectively, may be written as

act (b, left (right (sum (pre (a, p), compo (pre (b, q), pre (c, r))))))
= act (b, left (compo (pre (b, q), pre (c, r))))
= act (b, compol (pre (b, q), pre (c, r)))
= compol (act (b, pre (b, q)), pre (c, r))
= compol (q, pre (c, r))

The second group includes all axioms involving the operation symbol ‘back’. It is used to move the ‘position pointer’ back to an ancestor node along some path of the tree of a process structure. Note that at this time the ‘pointer’ is pointing to the process Q. Thus, if we want to perform an additional action c in the above example, we have first to perform a ‘back’ operation to move the ‘pointer’ back in order to let it point to the process composition (Q | c.R). Then we use the ‘right’ operation to let it point to the process c.R. Only then it is possible to apply the operation ‘act’ to the prefix c. The result is then:

$$Q \mid c.R \xrightarrow{c} Q \mid R,$$

as the following calculation shows :

act (c, right (back (compol (q, pre (c, r)))))
= act (c, right (compo (back (q), back (pre (c, r)))))
= act (c, right (compo (q, pre (c, r))))
= act (c, compor (q, pre (c, r)))
= compor (q, act (c, pre (c, r)))
= compor (q, r)

The structure of a process is like a tree. The ‘left’ and ‘right’ operations are used to go down the tree from the root to some action node. The ‘back’ operation is used to go up the tree to the root (of the whole tree or of some sub-tree), in order to get from there to another action node.

The third group of axioms is used to show how the ‘act’ operation functions.

The data types Cons + Trans are more complicated than Simple Construct + Simple Process Algebra in the sense that the former introduces non-determinism by allowing Sum and Compo constructs. As a consequence, there is no more homomorphism (isomorphism) in the usual sense between seed algebras. In order to see that，consider the following two processes that are bisimular:

**Example** 3.3

Q = a.b.nil + b.a.nil + a.b.nil,     Q' = a.nil | b.nil

Let q and q' be the seeds corresponding to Q and Q', where

q = sum (pre (a, pre (b, nil)), sum (pre (b, pre (a, nil)), pre (a, pre (b, nil)))),

q' = compo (pre (a, nil), pre (b, nil)).

The loose seed algebras LA (Trans, q) and LA (Trans, q') have the following elements:

LA (Trans, q) = {q,   left (q), right (q), left (right (q)), right (right (q)),
act (a, left (q)), act (b, left (right (q))), act (a, right (right (q))),
act (b, act (a, left (q))), act (a, act (b, left (right (q)))) ,
act (b, act (a, right (right (q)))), ……}

LA (Trans, q') = {q',   left (q'), right (q'), act (a, left (q')), act (b, right (q')),
back (act (a, left (q'))), back (act (b, right (q'))),
right (back (act (a, left (q')))), left (back (act (b, right (q')))),
act (b, right (back (act (a, left (q'))))), act (a, left (back (act (b, right (q'))))),
back (act (b, right (back (act (a, left (q')))))),
back (act (a, left (back (act (b, right (q')))))), …… }

The dots at the end of each loose seed algebra illustrate that these two loose seed algebras are actually infinite, because for example we know from the axioms of appendix 1 that for any element x of sort **proc,** the element back (x) is defined.

After building the quotient, the seed algebras are:

A (Trans, q) = {q,   pre (a, y), sum (pre (b, x), pre (a, y)), pre (b, x), y, x, nil}

A (Trans, q') = {q',   compol (x, y), compor (x, y), compol (nil, y), compor (x, nil),
compo (nil, y), compo (x, nil), compor (nil, y), compol (x, nil),
compol (nil, nil), compor (nil, nil), compo (nil, nil)}

where x = pre (a, nil), y = pre (b, nil).

Unlike loose seed algebra, they have only finitely many elements.

By taking a closer look at A (Trans, q) and A (Trans, q') we may find the following problems. First, there is no one-one correspondence between the elements of these two seed algebras. Second, in each of the two seed algebras there are elements equivalent to other elements headed by the non-hidden function 'act'. There are also elements equivalent to elements headed by hidden functions like left, right and back. The former one represents completed transitions of processes, the latter represents non-deterministic choices. These two kinds of functions should not be treated in an equal way. Third, even if we only consider elements headed by non-hidden functions, these elements of the two seed algebras do not have a one-one correspondence.

The intuitional reason of the third problem is obvious. The process Q has two possibilities of making an ***a***-transition, while the process Q' has only one possibility of making the same transition.  The former corresponds to two elements {act (a, left (q)), act (a, right (right (q)))} of the seed algebra of q, while the latter corresponds to only one element {act (a, left (q'))} of the seed algebra of q'. It is impossible to have a one-one correspondence between them. The idea is to establish the one-one correspondence on the set level. For each action ***x***, we let the set of all ***x***-transitions of Q correspond to the set of all ***x***-transitions of Q'. With the language of seed algebra, this idea may be stated as follows: For each non-hidden function f, we let the set of all elements of A (Trans, q) headed by f correspond to the set of all elements of A (Trans, q') headed by the same f. This idea leads to the definition of hidden extension and non-hidden closures.

**Definition** 3.1: Non-hidden closures and hidden extensions of a seed algebra A (D, ***a***), where ***a*** is a defined element of IN (D), are defined recursively as follows:

1. The seed ***a*** of the seed algebra A (D, ***a***) is a non-hidden closure of sort **s**, if ***a*** as an element of A (D, ***a***) is of sort s;

2. If $\alpha$ is a set of defined elements of sort **s**, then [ $\alpha$ ] denotes its hidden extension and is of sort **s**, where

4.1 If b $\in$ $\alpha$ , then b $\in$ [ $\alpha$ ]

4.2 If c $\in$ [ $\alpha$ ], f is hidden, f (c) is a defined element of A (D, a) and is of sort **s**, then f (c) $\in$ [ $\alpha$ ],

3. If g: $\mathbf{s_1 \times s_2 \times \ldots \times s_n \rightarrow s_{n+1}}$

is a non-hidden function of A (D, a). Given n element sets $\{x_1,...,x_n\}$ of sort $\mathbf{s_i}$, $0 \le i \le n$, where some or all of $\{x_1,...,x_n\}$ may be hidden extensions. Then $\overline{g}(x_1,...,x_n)$ = { $g(a_1,...,a_k)$ | $\forall i,\ a_i \in x_i \subseteq A_{s_i}$, $g(a_1,...,a_n)$ is a defined element of A (D, ***a***)} is a non-hidden closure (of $\{x_1,...,x_n\}$) of sort $\mathbf{s_{n+1}}$, provided that it is not empty, where $A_{s_i}$ is the set of elements of sort $\mathbf{s_i}$ of the initial algebra IN (D) of D, $n \ge 0$. $A_{s_i}$ is the carrier set of elements of sort s of IN (D). In case that all $x_i$ are single element we also write $\overline{g}(\{a_1\},...,\{a_n\}) = g(a_1,...,a_n)$ and $\overline{g} = g$ when n = 0.

Note that all hidden extensions and non hidden closures contain only defined elements. Note also that in some cases a non-hidden closure or hidden extension may contain infinitely many elements. But each element itself is finite. That means each element is only a finite nesting of function applications. Note also that the meaning of the notation [ $\alpha$ ] is context sensitive. In particular we should differentiate between isolate [ $\alpha$ ] and [ $\alpha$ ] as parameter of $\overline{f}$ , where f is some non-hidden function. The latter is a subset of the former because all elements of the later should satisfy the additional context constraints caused by the function f. Note further that sometimes a non-hidden closure and its hidden extension may denote the same thing. For example, assume $\alpha$ is a non-hidden closure and [ $\alpha$ ] its hidden extension. If there is no hidden function f, such that f (a) is a defined element of the seed algebra for some a $\in$ $\alpha$ , then [ $\alpha$ ] denotes nothing else than $\alpha$ itself.

In the following, we will use Greek letters $\alpha, \beta, \ldots$ to denote non-hidden closures and $[\alpha], [\beta], \ldots$ .to denote hidden extensions.

q = sum (pre (a, pre (b, nil)), sum (pre (b, pre (a, nil)), pre (a, pre (b, nil)))),

Consider example 3.3, [q] = {q, left (q) = right (right (q)), right (q), left (right (q))} = {q, pre (a, pre (b, nil)), sum (pre (b, pre (a, nil)), pre (a, pre (b, nil))), pre (b, pre (a, nil))} is the hidden extension of q. $\overline{act}$ ([a], [q]) = {act (a, left (q)) = act (a, right (right (q)))} = {pre (b, nil)} is a non-hidden closure. Here we see that the isolate [q] represents a four element set, but the [q] as parameter of the function act represents only a subset of the former, namely a one element set { pre (a, pre (b, nil)) }. This shows the context dependency of [q] as parameter of the function act. Furthermore, [$\overline{act}$ ([a], [q])] is the hidden extension of $\overline{act}$ ([a], [q]), containing the same one element set as $\overline{act}$ ([a], [q]) contains. On the other hand, $\overline{act}$ ([b], [$\overline{act}$ ([a], [q])]) = {nil} is a non-hidden closure. The case of q' is similar.

**Definition 3.2:** Let A (D, ***a***) and A (D, ***b***) be two seed algebras of the same data type D, where the seeds a, b are of the same sort. A mapping from the non-hidden closures of A (D, ***a***) to the non-hidden closures of A (D, ***b***), denoted with $h_{nh}$, is called a rough seed homomorphism from A (D, ***a***) to A (D, ***b***), where

1. $h_{nh}$ (***a***) = ***b***;

#According to definition 3.1, all seeds are non-hidden closures#

2. Let f: $s_1 \times s_2 \times \ldots \times s_k \rightarrow s_{k+1}$

be a non-hidden operation of D. For each non-hidden closure $\overline{f}([\alpha_1],...,[\alpha_n])$ of A (D, ***a***), where all i, $1 \le i \le n$, $\alpha_i$ are non-hidden closures and $[\alpha_i]$ their hidden extensions, there is a non-hidden closure $\overline{f}([\beta_1],...,[\beta_n])$ of A (D, ***b***), such that

$$h_{nh}(\overline{f}([\alpha_1],...,[\alpha_n])) = \overline{f}([\beta_1],...,[\beta_n]), \quad (3.1)$$

and $\forall$ i, $1 \le i \le n$, $h_{nh}(\alpha_i) = \beta_i$

3. $h_{nh}$ is called a rough seed isomorphism if there is a reversed rough seed homomorphism $h_{nh}^{-1}$ satisfying:

$\forall$ i, $h^{-1}_{nh}(\beta_i) = \alpha_i$,

$$h_{nh}^{-1}(h_{nh}(\overline{f}([\alpha_1],...,[\alpha_n]))) = h_{nh}^{-1}(\overline{f}([\beta_1],...,[\beta_n])) = \overline{f}([\alpha_1],...,[\alpha_n]), \quad (3.2)$$

We use the notation nhc (A (D, ***a***)) to denote the set of non-hidden closures of the seed algebra A (D, ***a***). A rough seed homomorphism (isomorphism) from A (D, ***a***) to A (D, ***b***) is also called a homomorphism (isomorphism) from nhc (A (D, ***a***)) to nhc (A (D, ***b***)).

We define a binary relation RSI on the set of non-hidden closures as follows: (nhc (A (D,

*a*)), nhc (A (D, *b*))) $\in$ RSI if and only if there is a rough seed isomorphism

$h_{nh}$: nhc (A (D, *a*)) → nhc (A (D, *b*)).

**Proposition 3.1:** RSI is an equivalence relation.

Proof: straightforward.

□

**Proposition 3.2:** The seed algebras A (Trans, q) and A (Trans, q') given in example 3.3 are rough seed isomorphic.

Proof: Reconsider the above example. With the concept of non-hidden closure, we rewrite the two seed algebras in the following way, where nhc (SA) means the set of all non-hidden closures of SA for any seed algebra SA:

nhc (A(Trans, q)) = {q, $\overline{act}$([a], [q]), $\overline{act}$([b], [q]), $\overline{act}$ ([b], [$\overline{act}$ ([a], [q])]), $\overline{act}$ ([a], [$\overline{act}$ ([b], [q])])}

nhc (A(Trans, q')) = {q', $\overline{act}$([a], [q']), $\overline{act}$([b], [q']), $\overline{act}$ ([b], [$\overline{act}$ ([a], [q'])]), $\overline{act}$ ([a], [$\overline{act}$ ([b], [q'])])}

This correspondence induces the following mapping:

$h_{nh}$ (q) = q'; for all x, $h_{nh}$ ($\overline{act}$ ([x], [q])) = $\overline{act}$ ([x], [q']),

for all x and y, $h_{nh}$ ($\overline{act}$ ([x], [$\overline{act}$ ([y], [q])])) = $\overline{act}$ ([x], [$\overline{act}$ ([y], [q'])])

It is easy to check that $h_{nh}$ is a rough seed isomorphism, which maps non-hidden closures to non-hidden closures in a bijective way.

□

## $ 4 Deep isomorphism and bisimulation

In this section we will discuss and compare isomorphism of seed algebras at two different levels. One is the rough seed isomorphism level as it was introduced in last section. The other one is the deep isomorphism introduced in this section. We will see that for strong bisimulation the rough seed isomorphism of seed algebras corresponds to trace equivalence of processes, while the deep isomorphism corresponds to strong bisimulation of processes. As result we will get the following conclusion: two processes are (strong) bisimular if and only if their corresponding seed algebras are deep (deep rough seed) isomorphic to each other.

Consider the following grammar of process algebra, where rp means the recursive call of a procedure. This process algebra includes almost all important aspects of CCS-like process algebra. Theoretically, we could refer to a complete CCS grammar. It is not difficult to prove theorem 4.1 based on the complete CCS grammar. But doing this way would introduce many unnecessary details. For our purpose, the following grammar is enough:

P ::= nil | $\tau$.P | y?x.P | y!x.P | P + Q | (P |Q) | rp (4.1)

x ::= a | b | u

y ::= c | d

Assume the transition rules are:

Input: $c?u.P \xrightarrow{ca} P[a/u]$ (input a name ***a*** via the channel ***c*** and put it in the place ***u***)

Output: $c!a.P \xrightarrow{c!a} P$ (output the name ***a*** via the channel ***c***)

Tau: $\tau.P \xrightarrow{\tau} P$ (invisible action)

Recur : $\dfrac{P \xrightarrow{\alpha} P'}{rp \xrightarrow{\alpha} P'}$, $rp =_{def} P$ (recursive call)

Sum1: $\dfrac{P \xrightarrow{\alpha} P'}{P + Q \xrightarrow{\alpha} P'}$ (non-deterministic choice 1)

Sum2: $\dfrac{Q \xrightarrow{\alpha} Q'}{P + Q \xrightarrow{\alpha} Q'}$ (non-deterministic choice 2)

Compo1: $\dfrac{P \xrightarrow{\alpha} P'}{P \mid Q \xrightarrow{\alpha} P' \mid Q}$ (parallel composition 1)

Compo2: $\dfrac{Q \xrightarrow{\alpha} Q'}{P \mid Q \xrightarrow{\alpha} P \mid Q'}$ (parallel composition 2)

Commu: $\dfrac{P \xrightarrow{c!a} P', Q \xrightarrow{ca} Q'}{P \mid Q \xrightarrow{\tau} P' \mid Q'}$ (communication)

Generally, there should be an additional condition $bv(\alpha) \cap fv(Q) = empty$ attached to the rules Compo1 and Compo2. But in our case, we do not need to include it because of two reasons. The first reason is that we are using an early transition semantics, rather than late semantics. The counter example given in [41] is based on late semantics where they use the rule $x(z).P \xrightarrow{x(z)}_l P$ for describing semantics of input transition. This would make the rules Compo1 and Compo2 invalid without the above condition. The second reason is that the syntax of our process grammar is not yet $\pi$ calculus syntax. Some important syntactic elements of $\pi$ calculus are yet missing. For example, the restriction operator $\nu$ x (also simply written as (x)) is missing. Therefore, the counter example given in [46] does not do any harm to the rules Compo1 and Compo2.

To give an algebraic semantics to this process algebra, we introduce the following ADT:

**Type** Process-Cons =

{**sort name, chan, proc, trace, ptrace**

**op**

a, b : → **name**

c, d : → **chan**

(): → **trace**

nil: → **proc**

**private op**
tau: **proc** → **proc**
i-proc : **chan** × **name** × **proc** → **proc**
o-proc : **chan** × **name** × **proc** → **proc**
sum: **proc** × **proc** → **proc**
compo: **proc** × **proc** → **proc**
rp: → **proc**
tconf: **proc** × **proc** × **trace** → **ptrace**
}
**End of Type** Process-Cons

**Type** Process-Trans =
{ Process-Cons +
**op**
tinput: **ptrace** × **chan** × **name** → **ptrace**
toutput: **ptrace** × **chan** × **name** → **ptrace**
t-h-act: **ptrace** → **ptrace**
**hop**
tcall: **ptrace** → **ptrace**
call: **proc** × **proc** → **proc**
it: **chan** × **name** → **trace**
ot: **chan** × **name** → **trace**
run: **trace** × **trace** → **trace**
input: **proc** × **chan** × **name** → **proc**
output: **proc** × **chan** × **name** → **proc**
h-act: **proc** → **proc**
subst: **proc** × **name** × **name** → **proc**

tleft : **ptrace** → **ptrace**
tright: **ptrace** → **ptrace**
tback: **ptrace** → **ptrace**

left : **proc** → **proc**
right: **proc** → **proc**
compol: **proc** × **proc** → **proc**
compor: **proc** × **proc** → **proc**
back: **proc** → **proc**
**axiom**
…………………………}

In order to avoid too many details, we leave the axiom part to Appendix 1.

The reader may have noticed that we have introduced a new operation ‘tconf’ in this section. The operation tconf is used to form a seed, which is a combination of sorts: **process** × **process** × **trace**. Here there are two things that are new.

First, it includes a sort **trace**, which is necessary because the bisimulation of two processes requires not only that they perform the same type of action (input/output), but also that they input or output the same name with the same channel. The sequence of (channel, name) pairs (, where it means input and ot means output) constitutes the input/output trace of a process (i.e. history of performing input/output operations). The tau actions are not observable and are thus not included in the trace.

Second, the **sort 'process'** appears twice in the definition of tconf. The first **process** pattern plays the same role as it has played in last section. It denotes the process undergoing transition (in terminology of process algebra) or function application (in terminology of seed algebra). The second **process** pattern is used to 'save' the original pattern of the process before any transition (function application) happens. The reason is that we need to keep a complete pattern of the process, which remains unchanged under any function application. This pattern of process will be used to restore the original process when there is a recursive call of procedure (note the **op** rp in the above ADT). This is the trick we use for processing recursive calls.

The recursive procedure call will be processed as a hidden operation since it is implicit in the definition of bisimulation. In order to simplify the discussion, we only treat the simple case that the recursive calls (denoted with the notation rp) do not include parameters. It is not difficult to include the more complicated case of recursive procedure calls with parameters. But the cost would be the introduction of many uninteresting details in the axiom system.

**Lemma 4.1**: There is a bijective correspondence between the processes generated by the process grammar (4.1) and the elements of the set $A_{\mathbf{proc}}$ of the ground term algebra of Process-Cons, where $A_{\mathbf{proc}}$ is the carrier set of all elements of sort **proc** in the seed algebra, such that
1. Each grammar rule r of (4.1) with P as the only non-terminal symbol on the left side corresponds to an operation o with result sort **proc** in Process-Cons, and vice versa;
2. The terminal symbols nil and rp correspond to the constant operations nil and rp respectively;
3. If grammatical terms $g_1, \ldots, g_n$ generated by grammar 4.1 correspond to algebraic terms $t_1, \ldots t_n$ generated by Process-Cons respectively, then each term g generated from $g_1, \ldots, g_n$ by rule r of grammar 4.1 corresponds to a term t generated from $t_1, \ldots t_n$ by operation o of Process-Cons, and vice versa.

Proof: The one-one correspondence between processes generated by the process grammar (xxx) and elements of $A_{\mathbf{proc}}$ can be established by a mapping

Ptop: {All processes generated by the non-terminal P} → $A_{\mathbf{proc}}$

as follows:

Ptop (nil) = nil,

Ptop (rp) = rp,

if Ptop (P) = p and Ptop (Q) = q then

{Ptop ($\tau$ .P) = tau (p)

Ptop (y?x .P) = i-proc (y, x, p),

Ptop (y!x .P) = o-proc (y, x, p)

Ptop (P + Q) = sum (p, q)

Ptop (P | Q) = compo (p, q)}

It is easy to check that the mapping Ptop can be reversed to a one-one mapping. The operations of the data type Process-Cons assure the completeness of process construction according to the original process grammar (4.1). Note that this correspondence is a pure syntactic affair. As for the semantic aspects, we will see that the operational semantics of the process grammar listed below corresponds to the axiom system of the data type Process-Trans listed in the Appendix.

□

Based on the result of this lemma we introduce the terminology 'P in process grammar notation' and 'p in element of seed algebra notation' to denote the two different representations of the same process in process grammar and in seed algebra. In general we have:

Ptop (P in process grammar notation) = p in element of seed algebra notation.

**Definition** 4.1: A seed in form of tconf (p, p, ()), where p is the seed algebra notation of some process P, is called an I-seed. Here the letter 'I' reminds of the fact that p is in its initial state with the empty trace (). This type of seed algebra is also called I-seed algebra.

In order to define deep isomorphism, we need seeds with non-empty traces. Those seeds are called S-seeds and will be introduced later.

**Definition 4.2:** In order to further simplify the notation of seed algebra, we introduce

$LA'(D, P)_{\text{P in process grammar notation}} =_{def} LA(D, tconf(p, p, ()))_{\text{p in element of seed algebra notation}}$

$A'(D, P)_{\text{P in process grammar notation}} =_{def} A(D, tconf(p, p, ()))_{\text{p in element of seed algebra notation}}$

to denote loose seed algebra and seed algebra, where () is the empty trace.

**Example 4.1:** 1. The notation A (D, tconf (sum (p, q), sum (p, q), ())), where sum (p, q) is element of seed algebra notation, can be simplified to A' (D, P+Q), where P+Q is in process grammar notation.

2. The notations A (Process-Trans, tconf (p, p, ())) and A (Process-Trans, tconf (q,q, ())), where p = sum (i-proc (c, u, o-proc (d, a, nil)), o-proc (d, a, i-proc (c, u, nil))), q = compo (i-proc (c, u, nil), o-proc (d, a, nil)), can be simplified to A' (Process-Trans, P) and A' (Process-Trans, Q), where P = c?u.d!a.nil + d!a.c?u.nil, Q = c?u.nil | d!a.nil.

We will check whether we can establish an algebraic semantics corresponding to the operational semantics of processes of the above grammar. In some cases this is true. See the following example.

**Example 4.2:** The seed algebras A' (Process-Trans, P) and A' (Process-Trans, Q) of example 4.1 are rough seed isomorphic to each other.

In fact, let S1 = tconf (p, p, ()), S2 = tconf (q, q, ()), it is

LA' (Process-Trans, P) = {S1, tleft (S1), tinput (tleft (S1), c, a), tinput (tleft (S1), c, b), tinput (tleft (S1), c, u), toutput (tinput (tleft (S1), c, a), d, a), toutput (tinput (tleft (S1), c, b), d, a), toutput (tinput (tleft (S1), c, u), d, a), tright (S1), toutput (tright (S1), d, a), tinput (toutput (tright (S1), d, a), c, a), tinput (toutput (tright (S1), d, a), c, b), tinput (toutput (tright (S1), d, a), c, u)},

LA' (Process-Trans, Q) = {S2, tleft (S2), tinput (tleft (S2), c, a), tinput (tleft (S2), c, b), tinput (tleft (S2), c, u), tback (tinput (tleft (S2), c, a)), tback (tinput (tleft (S2), c, b)), tback (tinput (tleft (S2), c, u)), tright (tback (tinput (tleft (S2), c, a))), tright (tback (tinput (tleft (S2), c, b))), tright (tback (tinput (tleft (S2), c, u))), toutput (tright (tback (tinput (tleft (S2), c, a))), d, a), toutput (tright (tback (tinput (tleft (S2), c, b))), d, a), toutput (tright (tback (tinput (tleft (S2), c, u))), d, a), tback (toutput (tright (tback (tinput (tleft (S2), c, a))), d, a)), tback (toutput (tright (tback (tinput (tleft (S2), c, b))), d, a)), tback (toutput (tright (tback (tinput (tleft (S2), c, u))), d, a)), tright (S2), toutput (tright (S2), d, a), tback (toutput (tright (S2), d, a)), tleft (tback (toutput (tright (S2), d, a))), tinput (tleft (tback (toutput (tright (S2), d, a))), c, a), tinput (tleft (tback (toutput (tright (S2), d, a))), c, b), tinput (tleft (tback (toutput (tright (S2), d, a))), c, u), tback (tinput (tleft (tback (toutput (tright (S2), d, a))), c, a)), tback (tinput (tleft (tback (toutput (tright (S2), d, a))), c, b)), tback (tinput (tleft (tback (toutput (tright (S2), d, a))), c, u))}

In order to show how to use equational reasoning of the axiom system to illustrate process transition, we take the element toutput (tinput (tleft (S1), c, b), d, a) as example:

toutput (tinput (tleft (S1), c, b), d, a) =

toutput (tinput (tleft (tconf (p, p, ())), c, b), d, a) =

toutput (tinput (tleft (tconf (sum (i-proc (c, u, o-proc (d, a, nil)), o-proc (d, a, i-proc (c, u, nil))), p, ())), c, b), d, a) =

toutput (tinput (tconf (i-proc (c, u, o-proc (d, a, nil)), p, ())), c, b), d, a) =

toutput (tconf (o-proc (d, a, nil)), p, run ((), it (c, b))), d, a) =

tconf (nil, p, run (run ((), it (c, b)), ot (d, a)))

The procedure of the above equational reasoning is at the same time the procedure of process transition. At each step of this equational reasoning, p is the original form of the process, nil is p after transition, the long expression run (run ((), it (c, b)), ot (d, a)) is the trace characterizing the transition from p to nil. This trace is the combination of the empty trace (), the input trace it (c, b) and the output trace ot (d, a). It is the seed algebraic notation of the trace produced by process transition.

It is easy to check that the following mapping $h_{nh}$ is a rough seed isomorphism:

$h_{nh}$ (S1) = S2,

$h_{nh}$ ($\overline{toutput}$ ([S1], [d], [a])) = $\overline{toutput}$ ([S2], [d], [a]),

For x = a, b, u:

{$h_{nh}$ ($\overline{tinput}$ ([S1], [c], [x])) = $\overline{tinput}$ ([S2], [c], [x]),

$h_{nh}$ ($\overline{toutput}$ ([$\overline{tinput}$ ([S1], [c], [x])], [d], [a])) = $\overline{toutput}$ ([$\overline{tinput}$ ([S2], [c], [x])], [d], [a]),

$h_{nh}$ ($\overline{tinput}$ ([$\overline{toutput}$ ([S1], [d], [a])], [c], [x])) = $\overline{tinput}$ ([$\overline{toutput}$ ([S2], [d], [a])], [c], [x])}

□

**Example** 4.3: Consider the seed algebras constructed with Process-Trans as ADT and S1 = tconf (p, p, ()), S2 = tconf (q, q, ()) as seeds respectively, where p and q are seed algebraic process notations of P $=_{def}$ d!a.P | d!b.P and Q $=_{def}$ d!a.Q + d!b.Q.

We don't want to list the elements of their seed algebras as in example 4.2. The new ingredient of this example is the recursion. It produces a long and even infinite list of elements.

The rough seed isomorphism is the following:
$h_{nh}$ (S1) = S2,

For any $\alpha, \beta$, if $h_{nh}$ ($\alpha$) = $\beta$ then for x = a, b:

{$h_{nh}$ ($\overline{toutput}$ ([$\alpha$], [d], [x])) = $\overline{toutput}$ ([$\beta$], [d], [x]) }

With the two examples given above one might think that the rough seed isomorphism of their corresponding seed algebras can be considered as a sufficient and necessary condition for the bisimulation of processes. But this is not true.

**Example** 4.4: Consider the following processes:

P = c!u.d!a.nil + c!u.d!b.nil,        Q = c!u.(d!a.nil + d!b.nil)

They are not bisimular, since the process P has no choice of outputting ***a*** or ***b*** after outputting ***u***. If the left summand is chosen, it can only output ***a***, otherwise it can only output ***b***. But Q still has a choice of either outputting ***a*** or outputting ***b*** after outputting ***u***.

However, it is still possible to design a rough seed isomorphism between the seed algebras A (Process-Trans, tconf (p, p, ())) and A (Process-Trans, tconf (q,q, ())) corresponding to these two processes. In order to prove that, let S1 = tconf (p, p, ()), S2 = tconf (q, q, ()) denote the two seeds, where

p = sum (o-proc (c, u, o-proc (d, a, nil)), o-proc (c, u, o-proc (d, b, nil)))
q = o-proc (c, u, sum (o-proc (d, a, nil), o-proc (d, b, nil)))

We calculate the hidden extensions and non-hidden closures iteratively.

[S1] = {S1, tleft (S1), tright (S1)},    [S2] = {S2},

$\overline{toutput}$ ([S1], [c], [u]) = {toutput (tleft (S1), c, u), toutput (tright (S1), c, u)},

$\overline{toutput}$ ([S2], [c], [u]) = {toutput (S2, c, u)},

[$\overline{toutput}$ ([S1], [c], [u])] = {toutput (tleft (S1), c, u), toutput (tright (S1), c, u)},

[$\overline{toutput}$ ([S2], [c], [u])] = {tleft (toutput (S2, c, u)), tright (toutput (S2, c, u))},

$\overline{toutput}$ ( [$\overline{toutput}$ ([S1], [c], [u])], [d], [a]) = {toutput (toutput (tleft (S1), c, u), d, a)},

$\overline{toutput}$ ( [$\overline{toutput}$ ([S1], [c], [u])], [d], [b]) = {toutput (toutput (tright (S1), c, u), d, b)}

$\overline{toutput}$ ( [$\overline{toutput}$ ([S2], [c], [u])] , [d], [a]) = {toutput (tleft (toutput (S2, c, u)), d, a)},

$\overline{toutput}$ ( [$\overline{toutput}$ ([S2], [c], [u])] , [d], [b]) = {toutput (tright (toutput (S2, c, u)) , d, b)}

We have the following mapping $h_{nh}$:

$h_{nh}$ (S1) = S2,

$h_{nh}$ ($\overline{toutput}$ ([S1], [c], [u])) = $\overline{toutput}$ ([S2], [c], [u]),

For x = a, b:

$h_{nh}$ ($\overline{toutput}$ ([$\overline{toutput}$ ([S1], [c], [u])], [d], [x])) =

$\overline{toutput}$ ([$\overline{toutput}$ ([S2], [c], [u])], [d], [x]),

That means that the existence of a rough seed isomorphism between seed algebras does not imply a bisimulation between the two processes they are corresponding to. What we get is only trace equivalence of processes.

**Definition 4.3:** Two processes P and Q are trace equivalent if for any action sequence $\mu_1\mu_2...\mu_n$ and transition sequence:

$$P \xrightarrow{\mu_1 \cdots \mu_n} P'$$

there is a transition sequence of Q:

$$Q \xrightarrow{\mu_1 \cdots \mu_n} Q'$$

and vice versa, where we use Greek letters $\mu, \nu, ...$ to represent actions like c?a, c!b, etc.

**Proposition 4.1:** Two processes are trace equivalent if and only if their corresponding I-seed algebras are rough seed isomorphic.

Proof: Given two trace equivalent processes P, Q and their seed algebras A' (D, P), A' (D, Q). Assume

$$\alpha = \overline{f_n}([\overline{f_{n-1}}(...\overline{f_1}(tconf(p, p, tr)...)...)]...) \in \text{nhc (A' (D, P))},$$

where p is the seed algebraic form of P and tr is any trace. It corresponds to the transition sequence:

$$(P, tr) \xrightarrow{\mu_1\mu_2 \cdots \mu_n} (P', tr')$$

where $\mu_i \in \{c?u, c!a, \tau\}$ are notations of process action for $f_i$. By definition of trace equivalence, there is a transition sequence

$$(Q, tr) \xrightarrow{\mu_1\mu_2 \cdots \mu_n} (Q', tr')$$

which corresponds to

$$\beta = \overline{f_n}([\overline{f_{n-1}}(...\overline{f_1}(tconf(q, q, tr)...)...)]...) \in \text{nhc (A' (D, Q))},$$

In this way we get a rough seed isomorphism

$h_{nh}$: nhc (A' (D, P)) → nhc (A' (D, Q))

with $h_{nh}$ ($\alpha$) = $\beta$. It is really a rough seed isomorphism since it satisfies all conditions of definition 3.2. This shows the truth of the 'only if' part of the proposition. Similarly we can prove the 'if' part.

□

This proposition explains why rough seed isomorphism of seed algebras is not equivalent to bisimulation of two processes. The answer is that the rough seed isomorphism of seed algebras is equivalent to trace equivalence of their corresponding processes, which is known to be weaker than bisimulation. We need a more powerful condition: in order to assure a bisimulation of two processes, not only their seed algebras should be rough seed isomorphic, but also their corresponding sub-seed algebras should be rough seed isomorphic.

In order to explain what are ‘corresponding sub-seed algebras’, we have to introduce the concept of corresponding pairs.

**Definition 4.4:** Given two seed algebras A (D, a) and A (D, b), which are rough seed isomorphic. Assume elements c $\in$ A (D, a), d $\in$ A (D, b). (c, d) is called a corresponding pair if and only if there are non-hidden closures $\alpha$ of nhc (A (D, a)) and $\beta$ of nhc (A (D, a)), such that $h_{nh}$ ( $\alpha$ ) = $\beta$ and c $\in \alpha$ and d $\in \beta$, respectively.

Now that we have defined the concept of a corresponding pair, a natural idea is to define the above mentioned “corresponding sub-seed algebras” as seed algebras with the corresponding pairs as seed. However, the elements of a corresponding pair usually do not take the form of a seed in a strict way, because they have already undergone transitions (non-hideen function applications) such that they usually don’t take the form tconf (p, p, ()) anymore. This means that the concept of seed algebra should be made more precise. At this place we will introduce another form of seed – the S-seed and S-seed algebra. We will then see that the “corresponding sub-seed algebras” should be S-seed algebras instead of I-seed algebras.

**Definition** 4.5: A seed in form of tconf (q, p, tr) is called a S-seed, when there exists an I-seed tconf (p, p, ()) and a sequence of function applications such that

$$f_n(g_{n,k}(g_{n,k-1}(...g_{n,1}(f_{n-1}(g_{n-1,j}(g_{n-1,j-1}(...g_{n-1,1}(f_{n-2}(...f_1(g_{1,l}(g_{1.l-1}(...g_{1,1}(tconf(p,p,())...)...)$$

$$= \quad tconf(q,p,tr) \qquad (4.2)$$

where all $f_i$ are non-hidden functions and all $g_{s,t}$ hidden functions, with $k,j,l,... \geq 0$.

Let P, Q be the process grammar notation of p and q. The equation (4.2) means that there is a transition sequence

$$P \xrightarrow{\mu_1} P_1 \xrightarrow{\mu_2} P_2 \xrightarrow{\mu_3} ... \xrightarrow{\mu_n} P_n = Q$$

such that each action $\mu_i$ corresponds to a non-hidden function $f_i$ in the way:

$P_j \xrightarrow{x!y} P_{j+1}$ ←→ toutput (tconf ($p_j$ , p, $tr_j$), x, y) = tconf ($p_{j+1}$ , p, run ($tr_j$, ot (x, y)))

$P_j \xrightarrow{x?y} P_{j+1}$ ←→ tinput (tconf ($p_j$ , p, $tr_j$), x, y) = tconf ($p_{j+1}$ , p, run ($tr_j$, it (x, y)))

$P_j \xrightarrow{\tau} P_{j+1}$ ←→ t-h-act (tconf ($p_j$ , p, $tr_j$)) = tconf ($p_{j+1}$ , p, $tr_j$)

Here the letter S reminds of the word sprout. The intuition is that the seed has become a sprout (after some transitions).

One may conclude from example 4.4 that not every corresponding pair can be used as seeds of “corresponding sub-seed algebras” (considered as S-seed algebras). But things are sometimes worse then one might expect. There are cases where none of the corresponding pairs can be used as seeds of “corresponding sub-seed algebras”.

**Example** 4.5: Consider example 4.4 once again. We investigate the rough seed isomorphism $h_{nh}$ from $N_q$ = nhc (A (Process-Trans, tconf (q, q, ()))) to $N_p$ = nhc (A (Process-Trans, tconf (p, p, ()))), where

p = sum (o-proc (c, u, o-proc (d, a, nil)), o-proc (c, u, o-proc (d, b, nil))),

q = o-proc (c, u, sum (o-proc (d, a, nil), o-proc (d, b, nil))).

$\alpha = \overline{toutput}$ ([S2], [c], [u]) $\in N_q$ and $h_{nh}$ ( $\alpha$ ) = $\beta = \overline{toutput}$ ([S1], [c], [u]) $\in N_p$ are two non-hidden closures, where S1 = tconf (p, p, ()), S2 = tconf (q, q, ()). In this case, there are only two corresponding pairs:

(p1 = toutput (tleft (S1), c, u), q1 = toutput (S2, c, u)),

(p2 = toutput (tright (S1), c, u), q1 = toutput (S2, c, u)).

Where p1, p2 $\in \alpha$ , q1 $\in \beta$ . For the first corresponding pair, the S-seed algebras are A (D, q1) and A (D, p1). We assert that there is no rough seed isomorphism $h'_{nh}$ from nhc (A (D, q1)) to nhc (A (D, p1)). For seeing that, consider $\overline{toutput}$ ([q1], [d], [b])) $\in$ nhc (A (D, q1)), which is well defined. Now there are two possibilities for defining $h'_{nh}$ ( $\overline{toutput}$ ([q1], [d], [b]))). Either it is equal to $\overline{toutput}$ ([p1], [d], [b])), which is however not defined. Or we can let $h'_{nh}$ ( $\overline{toutput}$ ([q1], [d], [b]))) = $\overline{toutput}$ ([p2], [d], [b])), which is not defined, too. This shows that there is no rough seed isomorphism between A (D, q1) and A (D, p1). Nor is there a rough seed isomorphism between A (D, q1) and A (D, p2).

□

The study on corresponding pairs leads us to the definition of deep (rough seed) isomorphism, which is given as follows:

**Definition 4.6:** Two (I- or S-) seed algebras A (D, a) and A (D, b) are called deep rough seed isomorphic (or simply: deep isomorphic) if and only if:

1. A (D, a) and A (D, b) are rough seed isomorphic by a mapping $h_{nh}$: nhc (A (D, a)) → nhc (A (D, b));
2. For each c $\in$ A (D, a), there is an element d $\in$ A (D, b), such that (c, d) is a corresponding pair, the S-seed algebras A (D, c) and A (D, d) are deep rough seed isomorphic.

The following proposition shows that the second condition of the above definition can be weakened.

**Proposition 4.2:**

Two (I- or S-) seed algebras A (D, **a**) and A (D, **b**) are deep isomorphic if and only if

1. A (D, **a**) and A (D, **b**) are rough seed isomorphic by the mapping $h_{nh}$: nhc (A (D, **a**)) → nhc (A (D, **b**));
2. For any element c $\in$ A (D, **a**) there is an element d $\in$ A (D, **b**), such that (c, d) is a corresponding

pair, and the seed algebras A (D, **c**) and A (D, **d**) are rough seed isomorphic with the same mapping $h_{nh}$.

Proof: First we prove that the condition presented in the statement of this proposition is necessary. That means if A (D, **a**) and A (D, **b**) fulfill the conditions in definition 4.6, then they also have the property stated in (the second part of) this proposition. That there is a rough seed homomorphism $h_{nh}$': nhc (A (D, **c**)) →nhc (A (D, **d**)) is trivial, since deep isomorphism implies rough seed isomorphism. What we have to do is to prove that $h_{nh}' = h_{nh}$, i.e. the new seed algebra A (D, **c**) is a sub-algebra of A (D, **a**) and A (D, **d**) a sub-algebra of A (D, **b**). Assume there are elements $\mathbf{u} \in$ A (D, **c**) and $\mathbf{v} \in$ A (D, **d**), such that (u, v) is a corresponding pair. According to the definition of seed algebra there are operation symbols $\{f_{ui}\}$, $\{g_{vj}\}$ of D, such that

$\mathbf{u} = f_{u1} (\ldots, f_{uk} (\ldots, \mathbf{c}, \ldots), \ldots)$

$\mathbf{v} = g_{v1} (\ldots, g_{vq} (\ldots, \mathbf{d}, \ldots), \ldots)$

Remember that we always assume that each operation has only one implementation. Therefore we can identify operations with functions. Again according to the definition of seed algebra there are operation symbols $\{f_{ci}\}$, $\{g_{dj}\}$ of D, such that

$\mathbf{c} = f_{c1} (\ldots, f_{cn} (\ldots, \mathbf{a}, \ldots), \ldots)$

$\mathbf{d} = g_{d1} (\ldots, g_{dm} (\ldots, \mathbf{b}, \ldots), \ldots)$

Combine the above equations, we have

$\mathbf{u} = f_{u1} (\ldots, f_{uk} (\ldots, f_{c1} (\ldots, f_{cn} (\ldots, \mathbf{a}, \ldots), \ldots), \ldots), \ldots)$

$\mathbf{v} = g_{v1} (\ldots, g_{vq} (\ldots, g_{d1} (\ldots, g_{dm} (\ldots, \mathbf{b}, \ldots), \ldots), \ldots), \ldots)$

This shows that $\mathbf{u} \in$ A (D, a) and $\mathbf{v} \in$ A (D, b), i.e.

A (D, **c**) $\subseteq$ A (D, **a**), A (D, **d**) $\subseteq$ A (D, **b**)

This shows that A (D, **c**) is a sub-algebra of A (D, **a**) and A (D, **d**) a sub-algebra of A (D, **b**). Therefore we may use the same rough seed isomorphism $h_{nh}$ for the mapping nhc (A (D, **c**)) → nhc (A (D, **d**)).

Now we prove that the condition presented in the statement of this proposition is also sufficient. If the second condition of definition 4.6 were not true, then either there is no corresponding pair (c, d) such that A (D, c) and A (D, d) are rough seed isomorphic, or for some element $u \in$ A (D, c), there is no $v \in$ A (D, d), such that (u, v) is a corresponding pair and A (D, u) und A (D, v) are deep isomorphic. Reasoning repeatedly in this way, we would get two chains of seed algebras. We use the notation **A < B** to denote that **A** is a (S-) seed algebra constructed with a seed from **B**. Then we would have …… < A (D, $c_n$) <A (D, $c_{n-1}$) <…<A (D, $c_1$) <A (D, $c_0$), and ……< A (D, $d_n$) <A (D, $d_{n-1}$) <…<A (D, $d_1$) <A (D, $d_0$), where $c = c_0$, $d = d_0$, such that for all $1 \leq i \leq n$, $c_i \in$ A (D, $c_{i-1}$) and $d_i \in$ A (D, $d_{i-1}$). Since A (D, $c_0$) and A (D, $d_0$) are known to be rough seed isomorphic, there are two possibilities. Either for all i, A (D, $c_i$) and A (D, $d_i$) are rough seed isomorphic, then the deep isomorphism is proved, against the assumption. Otherwise, there must be some n, A (D, $c_n$) and A (D, $d_n$) are not rough seed isomorphic. Use the same method as above we can prove that A (D, $c_n$) is a sub-seed algebra of A (D, $c_{n-1}$) and A (D, $d_n$) a sub-seed algebra of A (D, $d_{n-1}$). Thus they must be rough seed isomorphic. This refutes the assumption.

□

Remember the operational semantics of this process algebra we have given above, a definition of bisimulation is as follows:

**Definition 4.7:** A symmetric relation SBS on the set of all processes is called strong (early) bisimulation if for each (P, Q) $\in$ SBS, the following statement is true:

$\forall\ \mu,\ P \xrightarrow{\mu} P'$, there is Q' such that $Q \xrightarrow{\mu} Q'$

and (P', Q') $\in$ SBS, where $\mu = \tau$ or c?a or c!a for any channel c and any name a.

Two processes P and Q are called strongly bisimular (or: strongly bisimulate each other) if there is a strong bisimulation relation SBS such that (P, Q) $\in$ SBS.

In order to prove the following theorem, we have to differentiate between two kinds of transitions. A transition is called a concrete transition when the place where the transition occurs is explicitely given. Otherwise it is called an abstract transition.

**Example 4.6:** The transitions

$$c!a.P \xrightarrow{c!a} P, \tag{4.3}$$

$$c!b.P + c!a.Q \mid c!b.R \xrightarrow{c!a} Q \mid c!b.P \tag{4.4}$$

are all concrete transitions. Whereas

$$P \xrightarrow{c!a} P' \tag{4.5}$$

$$P \mid Q \xrightarrow{\tau} P' \mid Q' \tag{4.6}$$

$$P \xrightarrow{c!a} P, \quad \text{where } P =_{def} c!a.P + c!a.P \tag{4.7}$$

are all abstract transitions. In the last example above all details of the recursive process P are given. But still it is not possible to determine where the actual transition has occurred (the left c!a or right c!a)

In accordance with the two kinds of transitions, there are two kinds of correspondence. The first one is the corrspondence between concrete transitions and seed algebra elements whose outermost function is non-hidden. This corrspondence is obviously one-to-one. For example, based on the axiom system given in the Appendix, transition (4.3) correspondends to

toutput (tconf (o-proc (c, a, p), o-proc (c, a, p), ()), c, a) =
tconf (p, o-proc (c, a, p), run ((), ot (c, a)))

whereas transition (4.4) corresponds to

toutput (tleft ( tright (tconf (p', p', ()))), c, a) =
tconf (compo (q, o-proc (c, b, r)), p', run ((), ot (c, a)))

where

p' = sum (o-proc (c, b, p), compo (o-proc (c, a, q), o-proc (c, b, r))),

where p, q, r are the seed algebraic forms of P, Q and R, respectively.

But this kind of correspondence is less useful for our purpose, since any bisimulation requires the correspondence between abstract transitions of both processes. We need an algebraic concept, which can also denote abtract transitions because an abstract transition can represent all possible concrete transitions of the same kind (using the same channel to input/output the same name) while not explicitely mentioning them. This concept is nothing else than the non-hidden closure of seed algebra. This is the second correspondence mentioned above.

Thus transition (4.5) corresponds to

$\overline{toutput}$ ([ tconf (p, p, ())], [c], [a]) = $\overline{tconf}$ ({p'}, p, run ((), ot (c, a)))

where {p’} $=_{def}$ {p’ | p’ is the seed algebraic form of P’ such that $P \xrightarrow{c!a} P'$ }

Transition (4.6) corresponds to

$$\overline{t-h-act}\text{ ([ tconf (compo (p, q), compo (p, q), ())]) =}$$

$$\overline{tconf}\text{ (\{compo (p', q')\}, compo (p, q), ())}$$

where

{compo (p’, q’)} $=_{def}$ {r | r is the seed algebraic form of P’ | Q’ such that $P \mid Q \xrightarrow{\tau} P' \mid Q'$ }

Transition (4.7) corresponds to

$$\overline{toutput}\text{ ([ tconf (p, p, ())], [c], [a]) = }\overline{tconf}\text{ (p, p, run ((), ot (c, a)))}$$

**Theorem 4.1:** A pair of processes (P, Q) strongly bisimulate each other (i.e. process P is strongly bisimilar to process Q) iff their corresponding seed algebras A (D, a) and A (D, b) are deep isomorphic, where a = seed (tconf (p, p, ())), b = seed (tconf (q, q, ())). Note that p and q are the seed algebra element form of the processes P and Q.

Proof:

Since the proof involves lots of details, we sketch the main idea with following steps:

Step 1: Prove that there is a one-one correspondence between the processes generated by the process grammar (4.1) and the elements of the set $A_{\mathbf{proc}}$ of the ground term algebra of Process-Cons. This is done by lemma 4.1.

Step 2: Prove that there is a one-one correspondence between concrete transition sequences of a process P and elements of seed algebra with tconf (p, p, ()) as its seed, where p is the seed algebraic form of P, and the outermost function symbol of these elements are non-hidden function symbols. In order to do that, we have to use the concept of concrete and abstract transitions explained above. This can be done with structural induction.

Step 3: Prove that there is a one-one correspondence between abstract transition sequences of a process P and non-hidden closures of seed algebra with tconf (p, p, ()) as its seed.

As we said above, an abstract transition represents all possible concrete transitions of the same kind. Correspondingly, a non-hidden closure represents all seed algebra elements of the same kind (same outermost non-hidden function with same input/output channel and name). The correspondence between abstract transitions and non-hidden closures is also one-to-one. Since bisimulation is characterized with ability of doing abstract transitions, it is just this second kind of correspondence that we need for proving this theorem.

More exacty, the correspondence between abstract transitions of processes and non-hidden closures of seed algebra is established below, where the transition rules no more describe process to process transition, but transitions from process state to process state. Replace process by process state in transition rule is here only for convenience.

1. Each process P considered as zero transition corresponds to the seed tconf (p, p, ()), which is of the sort **ptrace**, where p is the algebraic form of P and () the zero trace;

2. The first transition of any process P

$$(P,()) \xrightarrow{\mu_1} (P',((),\mu_1)), \qquad \mu_1 = \text{ c?u or c!a}$$

or $$(P,()) \xrightarrow{\tau} (P',()),$$

corresponds to the non-hidden closure $\overline{f_1}([tconf(p,p,())],...)$, where $f_1$ is tinput, toutput or t-h-act, corresponding to the action c?u, c!a or $\tau$ that P may perform, and the dots …either represent a channel ***c*** and a name ***a*** in case of tinput and toutput, or represent nothing (empty) in case of invisible action $\tau$ .

3. If the transition sequence

$$(P,()) \xrightarrow{\mu_1\mu_2\cdots\mu_n} (P',tr), \qquad n \geq 1, \qquad \forall i, \mu_i \in \{\text{ c?u, c!a}, \tau \}$$

corresponds to the non-hidden closure $\overline{f_n}([\overline{f_{n-1}}(...[\overline{f_1}([tconf(p,p,())],...)]...)],...)$, where each $f_i$ is the non-hidden function corresponding to action $\mu_i$ and the dots … have the same meaning as above, $tr$ is the chain of non $\tau$ actions in $\mu_1\mu_2...\mu_n$. Then the transition sequences

$$(P,0) \xrightarrow{\mu_1\mu_2\cdots\mu_n\mu_{n+1}} (P',(tr,\mu_{n+1})) \ , \qquad \mu_{n+1} \in \{\text{ c?u, c!a }\}$$

$$(P,()) \xrightarrow{\mu_1\mu_2\cdots\mu_n\tau} (P',tr)$$

correspond to the non-hidden closure $\overline{f_{n+1}}([\overline{f_n}(...[\overline{f_1}([tconf(p,p,())],...)]...)],...)$, where $f_{n+1}$ is the non-hidden function representing the action $\mu_{n+1}$ or $\tau$, the dots … have the same meaning as above.

In this way, we have found a one-one correspondence between the abstract transition sequences (traces) of the process P and the non-hidden closures of the process states **ptrace** of the seed algebra with tconf (p, p, ()) as seed. At the same time we also found a one-one correspondence (4.8) between the non-hidden closures of the process states **ptrace** of two seed algebras with tconf (p, p, ()) respectively tconf (q, q, ()) as seeds if the two processes (strongly) bisimulate each other. More exactly, this is the one-one correspondence:

$$\overline{f_n}([\overline{f_{n-1}}(...\overline{f_1}([tconf(p,p,())],...)...)],...) \leftrightarrow$$

$$\overline{f_n}([\overline{f_{n-1}}(...\overline{f_1}([tconf(q,q,())],...)...)],...) \qquad (4.8)$$

where the corresponding dots on both sides either denote the same channel and name or are both empty.

Step 4: Prove that if P and Q are strongly bisimular, then A’ (D, P) and A’ (D, Q) are deep isomorphic.

From strong bisimilarity of P and Q their trace equivalence follows. By definition 3.2 the rough seed isomorphism of A’ (D, P) and A’ (D, Q) is guaranteed. Now assume

$$p' = f^{p}{}_{m}(f^{p}{}_{m-1} \dots f^{p}{}_{1}(tconf(p, p, ()) \dots) \dots) \in$$ A’ (D, P),

It corresponds to a concrete transition sequence:

$$(P, ()) \xrightarrow{\mu_1 \dots \mu_k}_{conc} (P', tr)$$

where $\mu_1, \dots, \mu_k$ corresponds to the non-hidden subsequence $f^{nh}{}_{1}, \dots, f^{nh}{}_{k}$ of the function sequence $f^{p}{}_{1}, \dots, f^{p}{}_{m}$.

By strong bisimularity there is a concrete transition sequence

$$(Q, ()) \xrightarrow{\mu_1 \dots \mu_k}_{conc} (Q', tr)$$

such that P’ and Q’ are strongly bisimular.

This trasition sequence corresponds to some element

$$q' = f^{q}{}_{n}(f^{q}{}_{n-1} \dots f^{q}{}_{1}(tconf(q, q, ()) \dots) \dots) \in$$ A’ (D, Q),

where $f^{q}{}_{1}, \dots, f^{q}{}_{n}$ has the same non-hidden subsequence $f^{nh}{}_{1}, \dots, f^{nh}{}_{k}$. Obviously (p’, q’) is a corresponding pair since they are contained in

$$\alpha = \overline{f^{nh}{}_{k}}(\dots[\overline{f^{nh}{}_{1}}([tconf(p, p, ())] \dots)] \dots)$$ and

$$\beta = \overline{f^{nh}{}_{k}}(\dots[\overline{f^{nh}{}_{1}}([tconf(q, q, ())] \dots)] \dots)$$

respectively. Now we know that p’ and q’ are seed algebraic form of P’ and Q’. Since P’ and Q’ are strongly bisimular, we can continue to apply the above prove procedure to p’ and q’ and start a new cycle of reasoning. By ignoring unimportant details, we conclude that to each $p'' = f^{p'}{}_{m'}(f^{p'}{}_{m'-1} \dots f^{p'}{}_{1}(tconf(p', p', ()))) \dots) \in$ A’ (D, P’) corresponds a transition

$$(P', ()) \xrightarrow{\mu_{1'} \dots \mu_{k'}}_{conc} (P'', tr')$$

which, by strong bisimulation, corresponds to another transition

$$(Q', ()) \xrightarrow{\mu_{1'} \dots \mu_{k'}}_{conc} (Q'', tr')$$

which, by assumption of this theorem, corresponds to the element:

$$q'' = f^{q'}{}_{n'}(f^{q'}{}_{n'-1} \dots f^{q'}{}_{1}(tconf(q', q', ()) \dots) \dots) \in$$ A’ (D, Q’)

This proves that A’ (D, P’) and A’ (D, Q’) are rough seed isomorphic. According to proposition 4.2 we conclude that A’ (D, P) and A’ (D, Q) are deep isomorphic

Step 5: Prove that if A’ (D, P) and A’ (D, Q) are deep isomorphic, then P and Q are strongly bisimular.

The proof of this step is almost the same as in step 4. We just need to reverse the proof procedure of step 4 to get the result.

□

## $5 Model Structure of Bisimulations

In the above two sections, we are limited to a two level structure (Process-Cons, Process-Trans) of ADT, which characterizes the strong early semantics. This is determined by the axiom system of Process-Trans. But if we want other type of semantics, we have to modify the axioms. More precisely, we will establish a directed acyclic graph of hierarchical ADT, each of whose nodes represents an ADT with a different axiom system, and thus also represents a different semantic model of bisimulation. The existence of a directed arc from node A to node B means that bisimulation of type A is weaker than that of type B. It follows then that the node of early strong bisimulation (discussed in previous sections) can not serve as source node of this graph. The source node of this directed acyclic graph is that node representing weak barbed bisimulation. There are only directed arcs going out from this node, but no arcs are going in. Let us start from the syntactic level:

First we recall our ADT of process grammar:

**Type** Process-Cons =
{**sort name, chan, proc, trace, ptrace**
**op**
a, b : → **name**
c, d : → **chan**
(): → **trace**
nil: → **proc**

**private op**
tau: **proc** → **proc**
i-proc : **chan** × **name** × **proc** → **proc**
o-proc : **chan** × **name** × **proc** → **proc**
sum: **proc** × **proc** → **proc**
compo: **proc** × **proc** → **proc**
rp: → **proc**
tconf: **proc** × **proc** × **trace** → **ptrace**
}
**End of Type** Process-Cons

In the following we will introduce step by step the variety of bisimulation semantics using deep isomorphism of seed algebras. We start from the least powerful bisimulation, the weak barbed bisimulation.

**Definition 5.1:** We introduce the notations: $\xrightarrow{*}$ means $(\xrightarrow{\tau})*$ ; $\xrightarrow{*\alpha*}$ means $(\xrightarrow{\tau})* \xrightarrow{\alpha} (\xrightarrow{\tau})*$. Read the former notation as 'zero or a finite number of $\tau$ actions', and the second notation as 'an $\alpha$ action preceded and followed by zero or a finite number

of $\tau$ actions'

**Definition** 5.2 **(Weak Barbed Bisimulation)** A symmetric binary relation WBB is called a weak barbed bisimulation if for each (P, Q) $\in$ WBB,

1. $P \downarrow_{c?}$ implies there is $Q'$ such that $Q \xrightarrow{*} Q'$ and $Q' \downarrow_{c?}$, for any channel c;
2. $P \downarrow_{c!}$ implies there is $Q'$ such that $Q \xrightarrow{*} Q'$ and $Q' \downarrow_{c!}$, for any channel c;
3. If $P \xrightarrow{\tau} P'$ then there is $Q'$ such that $Q \xrightarrow{*} Q'$ and $(P', Q') \in WBB$.

P and Q are called weakly barbed bisimilar, if there is a weak barbed bisimulation WBB such that $(P, Q) \in WBB$.

Before discussing the technical details, let us note that the keyword here is 'weak'. It is the source of all major difficulties we will be dealing with. From the point of view of process algebra, 'weak' means ignoring $\tau$ actions when deciding about the bisimulation of two processes. But in seed algebra, 'weak' means a miss of balance when deciding about rough seed isomorphism and deep isomorphism. By balance we mean the balance of actions. Note that in the case of WBB, there is no one-one correspondence between concrete transition sequences and elements of seed algebra, because we have general transitions such as $\xrightarrow{*}$ and even barb check, which is not a transition at all. Here we use a more general concept: gaction (generalized action).

Remember the proof of theorem 4.1 in section 4, where the two strong early bisimular processes perform the same (input or output or $\tau$ ) action. This shows that the actions of both processes are well balanced. As a consequence, their corresponding seed algebras have a perfect one-one correspondence of non-hidden closures. However, for weak bisimulation the things are different. While one process performs a $\tau$ action, the other process performs a gaction, which may be a $\tau$ action, a finite sequence of $\tau$ actions (a $\tau$ series for short), or no action at all. For convenience we keep using the notation 'action' instead of 'gaction' unless we feel we must emphasize the word 'gaction' in order to avoid ambiguity. This shows that the actions of both processes are not balanced. This fact reminds us of the necessity of differentiating the actions performed by two processes. If process P performs action a1 and process Q performs action b1 to simulate a1, then we say that a1 is the active action and b1 the passive action. The passive action must meet some conditions, but is not necessary equal to the active action. The difficulty is how to design non-hidden and hidden functions such that two weakly bisimulating processes can still be represented by two deep isomorphic seed algebras despite of the unbalance of active and passive actions.

In order to define barb checks, we introduce two non-hidden functions:

1. t-barb-i: check the availability of some input channel.

2. t-barb-o: check the availability of some output channel.

In order to describe unbalanced active and passive actions, we have to introduce another two new functions t-or-no, which is a non-hidden operation, and h-or-no, which is a hidden operation. The t-or-no operation, when applied to a process P, may perform a tau action or do nothing, depending on whether a tau action is feasible and desirable. The h-or-no operation, on the other hand, always performs a tau action. The t-or-no operation is used to construct non-hidden closures to implement unbalanced active and passive actions, while the repeating application of the hidden operation h-or-no is used to implement a finite sequence of $\tau$ actions to make active and passive actions even more unbalanced.

We propose the following ADT for weak barbed bisimulation. Note that the trace tr is here not used in the reasoning of axioms, because no input neither output happens. It remains empty in the whole process of barbed bisimulation. We need it only in bisimulation where either input or output happens.

**Type** WBB =

{ **Type** Process-Cons

+ **sort** **Bool**

+ **op**

t-barb-i: **ptrace** × **chan** → **ptrace**

t-barb-o: **ptrace** × **chan** → **ptrace**

t-or-no: **ptrace** → **ptrace**

+ **hop**

twrap: **ptrace** → **ptrace**

tleft: **ptrace** → **ptrace**

tright: **ptrace** → **ptrace**

tback: **ptrace** → **ptrace**

tcall: **ptrace** → **ptrace**

h-or-no: **ptrace** → **ptrace**

compol: **proc** × **proc** → **proc**

compor: **proc** × **proc** → **proc**

subst: **proc** × **name** × **name** → **proc**

wrap: **proc** → **proc**

h-act: **proc** → **proc**

or-no: **proc** → **proc**

barb-i: $\mathbf{proc} \times \mathbf{chan} \to \mathbf{proc}$

barb-o: $\mathbf{proc} \times \mathbf{chan} \to \mathbf{proc}$

\+ **axiom**

Abstract of axiom semantics:

1. The t-barb-i operation checks the existence of any input barb. The application of the function barb-i to process p is undefined if such a barb is not fund in p. Note that t-barb-i does not perform any $\tau -$ action to find such a barb. Instead we can use sufficiently many h-or-no functions to do that.

2. The t-barb-o operation has a similar action for output barb.

3. The t-or-no function may perform a $\tau$ action by locating its position using tleft or tright operation if necessary.

4. But t-or-no may also perform no action at all if it is preceded by a twrap operation, which wraps the original process with a nil-composition. In this way we accomplish that applying the same non-hidden operation t-or-no to two processes produces two different effects: while one process performs a $\tau$ action the other one performs nothing.

5. The hidden operation h-or-no is needed when we require that the simulating process Q has to perform more than one $\tau$ action. Each application of h-or-no performs one more $\tau$ action. As we said before, all hidden operation applications are ignored when we check the deep isomorphism of non-hidden closures.

The whole axiom system of WBB is listed in Appendix 2.

}

**End of Type** WBB

The following example shows the necessity of introducing the non-hidden function twrap:

**Example 5.1:** Consider the pair of processes: $P = \tau.(c!z \mid c?y)$, $Q = c!z \mid c?y$

Obviously they are weakly barbed bisimular. We let P perform a $\tau$ action and become $P' = c!z \mid c?y$, while Q does nothing. Now it is $P' \equiv Q$. They can do both a $\tau$ action. They also both have an input barb and an output barb and fulfill all requirements of weak barbed bisimulation. However, if we will be sure that applying t-or-no operation to Q does not produce any $\tau$ action, we have to first wrap Q in a nil-composition like compo (q, nil). In this way the application of t-or-no to Q will do no harm to it and keep it as it is, since

t-or-no (twrap (tconf (q, q, ()))) = t-or-no (tconf (wrap (q), q, ()))

= t-or-no (tconf (compo (q, nil), q, ())) = tconf (compo (q, nil), q, ())

**Example 5.2: (P =** $\tau.x!a + y!a$ **, Q =** $x!a + y!a$ **)** are not weak barbed bisimular.

Consider the non-closure $NC_1 = \overline{t-or-no}\ (tconf(p,p,()))$ = {tconf (p, p, ()), p'} there is a corresponding non-hidden closure $NC_2$ = $\overline{t-or-no}(tconf(q,q,()))$ = {tconf (q, q, ())}, where p' = tconf (o-proc (x, a, nil), p, ()). So far so good.

But if we select p' from $NC_1$, then there is only one element tconf (q, q, ()) from $NC_2$, which could be a corresponding element of p'. However, (A (WBB, p'), A (WBB, tconf (q, q, ()))) does not form a pair of corresponding sub-seed algebras, since $\overline{t-barb-o}([\,tconf(q,q,())\,]$, [y]) is defined, while $\overline{t-barb-o}$ ([p'], [y]) is not defined (more exactly, the set of defined elements contained in this non-hidden closure is empty).

**Theorem 5.1:** A pair of processes (P, Q) $\in WBB$ iff their corresponding seed algebras A (WBB, a) and A (WBB, b) are deep isomorphic, where a = tconf (p, p, ()), b = tconf (q, q, ()). Note that p and q are the seed algebra element form of the processes P and Q.

Proof: We have already seen in theorem 4.1 about how to prove such deep isomorphism results, where the proof procedure consists of five steps. Since the basic proof idea is the same, we will only focus on those points of the proof that are different from theorem 4.1. In the following we condense the five steps in four steps:

Step 1: Prove the one-one correspondence between the processes of the process grammar (4.1) and the elements of the **sort proc** in the ground term algebra of Process-Cons. This remains to be true here.

Step 2: Prove that if (P, Q) $\in WBB$ then A (WBB, a) and A (WBB, b) are rough seed isomorphic.

We prove this conclusion alone the way: nhc (A (WBB, a)) $\xrightarrow{1}$ active action (P) $\xrightarrow{2}$ passive action (Q) $\xrightarrow{3}$ nhc (A (WBB, b)). Note that for processes P and Q with (P, Q)$\in$WBB, there are only three kinds of active actions corresponding to seven kinds of passive actions. For a detailed representation of the above mapping $\xrightarrow{1,2,3}$ see table 5.1, where the mapping is to read from left to right. For example, the first item of this table is read as follows: a non-hidden closure $\overline{t-barb-i}$ of nhc (A (WBB, a)) (possibly with a locating sequence, where a locating sequence is a finite sequence of hidden locating functions tleft, tright, tback and tcall), corresponds uniquely to an input barb check (an active action), which corresponds uniquely to the same input barb check (a passive action, possibly with a $\tau$ series), which corresponds uniquely to a non-hidden closure $\overline{t-barb-i}$ of nhc (A (WBB, a)) (possibly with a locating sequence or even a h-or-no series) It is most important to note that no matter how diverse are the passive actions corresponding to a single active action, the axiom system of WBB is designed in the way such that the resulting mapping from nhc (A (WBB, a)) to nhc (A (WBB, b)) is unique. Moreover, each non-hidden closure $\overline{f}(...)$ with non-hidden function f is mapped to a non-hidden closure $\overline{f}(...)$ with the same function f, where the possible application of hidden functions is also shown. A h-or-no series is a finite sequence of h-or-no functions. It is easy to see that the reverse mapping is also unique. More exactly, we obtain the reverse mapping when we change the first line of table 5.1 to (nhc (A (WBB, b)), Active action, Passive action, nhc (A (WBB, a))) and let everything else unchanged. This proves the rough seed isomorphism of nhc (A (WBB, a)) and nhc (A (WBB, b)).

Step 3: Prove that if (P, Q) $\in WBB$ then A (WBB, a) and A (WBB, b) are deep isomorphic.

We use $P \xrightarrow{\beta_1 \ldots \beta_n} P'$ to denote that process P shows a sequence of action (more precise: gaction) $\beta_1, \beta_2, ..., \beta_n$ and then becomes P'.

Assume

$$p' = f^p{}_m(f^p{}_{m-1} \ldots f^p{}_1(tconf(p, p, ())...)...) \in \text{ A (WBB, a),}$$

It corresponds to a gaction sequence:

$$P \xrightarrow{\beta_1 \ldots \beta_k} P' \quad (5.1)$$

where $\beta_1, \beta_2, ..., \beta_k$ corresponds to the non-hidden subsequence $f^{nh}{}_1, ..., f^{nh}{}_k$ of the function sequence $f^p{}_1, ..., f^p{}_m$.

By weak barbed bisimularity there is a gaction sequence

$$Q \xrightarrow{\beta'_1 \ldots \beta'_k} Q' \quad (5.2)$$

such that P' and Q' are weak barbed bisimular.

This gaction sequence corresponds to some element

$$q' = f^q{}_n(f^q{}_{n-1} \ldots f^q{}_1(tconf(q, q, ())...)...) \in \text{ A (WBB, b),}$$

where $f^q{}_1, ..., f^q{}_n$ has the same non-hidden subsequence $f^{nh}{}_1, ..., f^{nh}{}_k$. Obviously (p', q') is a corresponding pair since they are contained in

$$\gamma = \overline{f^{nh}{}_k}(...[\overline{f^{nh}{}_1}([tconf(p, p, ())]...)]...) \quad \text{and}$$

$$\delta = \overline{f^{nh}{}_k}(...[\overline{f^{nh}{}_1}([tconf(q, q, ())]...)]...)$$

respectively. Since P' and Q' are weak barbed bisimular, we can continue to apply the above prove procedure to p' and q' and start a new cycle of reasoning. By ignoring unimportant details, we conclude that to each $p'' = f^{p'}{}_{m'}(f^{p'}{}_{m'-1} \ldots f^{p'}{}_1(tconf(p', p', ()))...) \in$ A (WBB, a'), where a' = tconf ( $p''$ , $p''$ , ()), corresponds a action sequence

$$P' \xrightarrow{\beta_{1'}\ldots\beta_{k'}} P''$$

which, by weak barbed bisimulation, corresponds to another action sequence

$$Q' \xrightarrow{\beta_{1'}\ldots\beta_{k'}} Q''$$

which, by assumption of this theorem, corresponds to the element:

$q''= f^{q'}{}_{n'}( f^{q'}{}_{n'-1}\ldots f^{q'}{}_{1}(tconf\,(q',q',())\ldots)\ldots) \in$ A (WBB, b')

Where b' = tconf ( $q'',q''$ , ()), This proves that A (WBB, a') and A (WBB, b') are rough seed isomorphic. According to proposition 4.2 we conclude that A (WBB, a) and A (WBB, b) are deep isomorphic

Step 4: Prove that if A (WBB, a) and A (WBB, b) are deep isomorphic then (P, Q) $\in WBB$ .

Reverse the proof procedure of step 2 and step 3 and we get the result.

□

Note that each gaction, no matter it is an active or passive action in table 5.1, corresponds to one and only one non-hidden closure. This coincides with the fact that the number of active actions $\beta_i$ in (5.1) equals to the number of passive actions $\beta'_i$ in (5.2).

Table 5.1 rough seed isomorphic mapping for WBB

| nhc (A (WBB, a)) | Active action | Passive action | nhc (A (WBB, b)) |
|---|---|---|---|
| Locating sequence + $\overline{t-barb-i}$ | input barb check | Same input barb check | Locating sequence + $\overline{t-barb-i}$ |
| | | $\tau$ series + Same input barb check | h-or-no series + Locating sequence + $\overline{t-barb-i}$ |
| Locating sequence + $\overline{t-barb-o}$ | output barb check | Same output barb check | Locating sequence + $\overline{t-barb-o}$ |
| | | $\tau$ series + Same output barb check | h-or-no series + Locating sequence + $\overline{t-barb-o}$ |
| Locating sequence + $\overline{t-or-no}$ | $\tau$ action | $\tau$ action | Locating sequence + $\overline{t-or-no}$ |
| | | $\tau$ series | h-or-no series + Locating sequence + |
| | | | $\overline{t-or-no}$ |
| | | no action | Twrap + $\overline{t-or-no}$ |

The proof of this theorem is fundamental for all other theorems in the remaining part of this section. The basic idea of the proofs of other theorems is almost the same as that given in this theorem. The core of all proofs is the table illustrating the rough seed isomorphism between seed algebras, like table 5.1 for theorem 5.1. Each different type of bisimulation is characterized by a different table of this kind, just as weak barbed bisimulation is characterized by table 5.1. Any thing else, including the proof of deep isomorphism itself, is almost the same for all bisimulations. Therefore, roughly speaking, switching to a new kind of bisimulation is equal to designing a new table for the proof of the rough seed isomorphism for its seed algebras.

**Definition** 5.3 **(Strong Barbed Bisimulation)** A symmetric binary relation SBB is called a strong barbed bisimulation if for each (P, Q) $\in$ SBB,

(1) $P\downarrow_{c?}$ implies $Q\downarrow_{c?}$, for any channel c;

(2) $P\downarrow_{c!}$ implies $Q\downarrow_{c!}$, for any channel c;

(3) If $P\xrightarrow{\tau}P'$ then there is $Q'$ such that $Q\xrightarrow{\tau}Q'$ and $(P',Q')\in SBB$.

P and Q are called strongly barbed bisimular, if there is a strong barbed bisimulation SBB such that $(P,Q)\in SBB$.

The strong barbed bisimulation SBB makes each $\tau$ action explicit.

**Type** SBB =

{ **Type** WBB

**+ op**

t-h-act: **ptrace → ptrace**

- hop

h-or-no

**+ axiom**

*# non-hidden Operation#*

t-h-act (tconf (q, p, tr)) = tconf (h-act (q), p, tr)

}

**End of Type** SBB

Abstract of axiom semantics:

1. Different from t-or-no, t-h-act performs one and only one $\tau$ action.
2. SBB does not inherit h-or-no function from WBB. This excludes the possibility of performing $\tau$ actions arbitrary.

There is only one new axiom in addition to axioms of WBB.

}

**End of Type** SBB

**Theorem 5.2:** A pair of processes (P, Q) are strong barbed bisimilar iff their corresponding seed algebras A (SBB, a) and A (SBB, b) are deep isomorphic, where a = tconf (p, p, ()), b = tconf (q, q, ()). Note that p and q are the seed algebra element form of the processes P and Q.

Proof: The proof idea of theorem 5.1 can be basically applied here. The case for strong bisimulation is much simpler to deal with than that for weak bisimulation. For the Only-if part, we show that nhc (A (SBB, a)) $\xrightarrow{1}$ active action (P) $\xrightarrow{2}$ passive action (Q) $\xrightarrow{3}$ nhc (A (SBB, b)) is a rough seed isomorphism between non-hidden closures of both seed algebras. See the table 5.2, which is much simpler than table 5.1. Note that the data type SBB inherits everything from data type WBB except the hidden function h-or-no. Correspondingly, the first three non-hidden closures of this table are inherited from table 5.1 with their behavior limited, such that the passive actions of input barb check and output barb check are now exact (any additional $\tau$ series is impossible). On the other hand, since the hidden function h-or-no is excluded from SBB, the non-hidden closures $\overline{t-barb-i}$ and $\overline{t-barb-o}$ now cannot be attached with any h-or-no series. Also the non-hidden closure $\overline{t-or-no}$ can no more be attached with any h-or-no series. But 'no action' is 'formally' still allowed. The word 'formally' means that although 'noaction' is allowed, it will be not chosen as an appropriate passive action when there is risk that the strong barbed bisimulation will be destroyed. See example 5.3. Thus the third item of table 5.2 has now two possible passive actions instead of three. All these guarantee the one-one correspondence between passive actions and non-hidden closures of nhc (A (SBB, b)).

For proving the If part, it is easy to reconstruct the proof chain to active action (P) $\xrightarrow{1}$ nhc (A (SBB, a)) $\xrightarrow{2}$ nhc (A (SBB, b)) $\xrightarrow{3}$ passive action (Q).

The idea of proving the deep isomorphism is then similar to the third step of the proof of theorem 5.1. We will not repeat it at this place.

□

**Example** 5.3 Let P = $\tau.\tau.nil$, Q = $\tau.\tau.nil$

No doubt it is $(P,Q) \in SBB$. There corresponding seed algebras are A (SBB, p) and A (D, SBB, q), where p = tau (tau (nil)), q = tau (tau (nil)).

It is harmless to apply WBB functions to them (except h-or-no, which is not inherited). Let the first step of non-hidden closure construction be:

$\overline{t-or-no}$([tconf (p, p, ())]) = {tconf (tau (nil), p, ())}

The problem is how to construct the non-hiddn closure for the passive action of q. The hidden extension of its seed is:

[tconf (q, q, ())] = {twrap (tconf (q, q, ())), tconf (q, q, ())}

But as we said above, the hidden extension is context sensitive. Thus

$\overline{t-or-no}$([tconf (q, q, ())]) = $\overline{t-or-no}$({tconf (q, q, ())})

= {tconf (tau (nil), p, ())}

This shows that the possible 'no action' of q is only formally allowed. In the practical construction of non-hidden closure, it will be excluded by the context. This is why keeping 'no action' in table 5.2 is harmless.

Table 5.2 rough seed isomorphic mapping for SBB

| nhc (A (SBB, a)) | Active action | Passive action | nhc (A (SBB, b)) |
|---|---|---|---|
| Locating sequence + $\overline{t-barb-i}$ | input barb check | Same input barb check | Locating sequence + $\overline{t-barb-i}$ |
| Locating sequence + $\overline{t-barb-o}$ | output barb check | Same output barb check | Locating sequence + $\overline{t-barb-o}$ |
| Locating sequence + $\overline{t-or-no}$ | $\tau$ action | $\tau$ action | Locating sequence + $\overline{t-or-no}$ |
| | | No action | Twrap + $\overline{t-or-no}$ |
| Locating sequence + $\overline{t-h-act}$ | $\tau$ action | $\tau$ action | Locating sequence + $\overline{t-h-act}$ |

**Theorem 5.3: F**rom the deep isomorphism of A (SBB, a) and A (SBB, b) it follows the deep isomorphism of A (WBB, a) and A (WBB, b), but not vice versa, where a, b have the same meaning as in theorem 5.2.

Proof: To be more exact, we restate the theorem as follows: if A (SBB, a) and A (SBB, b) are deep isomorphic with respect to all sorts, operations and axioms of the data type SBB, including those inherited from WBB, then A (WBB, a) and A (WBB, b) are deep isomorphic with respect to all sorts, operations and axioms of the data type WBB only. The difference is that in order to prove the former, one need to test all possible non-hidden closures constructed with non-hidden functions from SBB and from WBB, but in order to prove the latter, one need only to test non-hidden closures constructed with non-hidden functions from WBB. Proving the latter is only part of the job of proving the former. This can be easily seen when we note that the table 5.1 is part of the table 5.2, with the semantics of its non-hidden closures, its active and passive actions reduced to not include any h-or-no series ($\tau$ series). This is necessary because of the requirement of strong bisimularity. Therefore the conclusion of this theorem is already implied by theorem 5.2.

□

**Definition 5.4 (Weak Ground Bisimulation)** A symmetric binary relation WGB is called a weak ground bisimulation if for each $(P,Q)\in WGB$ , there is $z \notin fn(P,Q)$ such that

(1) if $P \xrightarrow{\alpha} P'$ where $\alpha$ is $c!x$ or $cz$ , there is $Q'$ such that $Q \xrightarrow{*\alpha*} Q'$ and $(P',Q')\in WGB$ .

(2) if $P \xrightarrow{\tau} P'$ then there is $Q'$ such that $Q \xrightarrow{*} Q'$ and $(P',Q')\in WGB$ .

P and Q are called weakly ground bisimilar, if there is a weak ground bisimulation WGB such that $(P,Q)\in WGB$ .

(Weak) ground bisimulation is stronger than (weak) barbed bisimulation, since it takes the non-tau transitions into consideration. In order to get the ground bisimulation model from the barbed one, we have to insert new operations and axioms for input/output transitions. The definition of weak ground bisimulation shows that it suffices to consider a single fresh name as the object of any input action.

**Type** WGB’ =

{ **Type** WBB

+ **op**

tinput: **ptrace** $\times$ **chan** $\times$ **name** $\rightarrow$ **ptrace**

toutput: **ptrace** $\times$ **chan** $\times$ **name** $\rightarrow$ **ptrace**

+ **hop**

it: **chan** $\times$ **name** $\rightarrow$ **trace**

ot: **chan** $\times$ **name** $\rightarrow$ **trace**

run: **trace** $\times$ **trace** $\rightarrow$ **trace**

input: **proc** $\times$ **chan** $\times$ **name** $\rightarrow$ **proc**

output: **proc** $\times$ **chan** $\times$ **name** $\rightarrow$ **proc**

+ **axiom**

**Abstract of axiom semantics:**

1. The semantics of input and output axioms have been explained in previous sections.
2. However, the capability of input operation is limited. The input axiom input (i-proc (c, x, p), c, x) = p allows only a particular fresh name x to be inputted. This corresponds to the requirement that in ground bisimulation only a single flash name can be considered for input.

The whole axiom system of WGB is listed in Appendix 3.

}

**End of Type** WGB’

In order to prove the correspondence between weak ground bisimulation and deep isomorphism of WGB’ seed algebras, we need a lemma (Note that we use the notation WGB in two senses. It is used both as the name of an ADT, like WGB’ above, and the name of a type of bisimulation, like WGB’ in the following lemma. This is also true for other bisimulation types):

**Lemma 5.1:** The first part of definition 5.4 can be modified to the following while the semantics of weak ground bisimulation is not changed:

(1)’ if $P \xrightarrow{\alpha} P'$ where $\alpha$ is $c!x$ or $cz$, there is $Q'$ such that $Q \xrightarrow{*\alpha} Q'$ and $(P', Q') \in WGB$.

Proof: For the moment we call the modified definition of weak ground bisimulation as WGB’. It is easy to see that if $(P,Q) \in WGB'$ then $(P,Q) \in WGB$. Now we prove the reverse conclusion. Given any WGB, we construct a new relation WGB’ as follows:

1. For all (P, Q), if $(P,Q) \in WGB$ then $(P,Q) \in WGB'$
2. If $(R,S) \in WGB$, $R \xrightarrow{\beta} P$, $S \xrightarrow{*\beta} Q \xrightarrow{*} S'$, $(P,S') \in WGB$, then $(P,Q) \in WGB'$;

We assert that WGB’ is a weak ground bisimulation according to the new definition. To be convinced about that, assume any $(P,Q) \in WGB'$:

1. If $P \xrightarrow{\alpha} P'$ where $\alpha$ is $c!x$ or $cz$, then

1.1 Either $(P,Q) \in WGB$, then we have $Q \xrightarrow{*\alpha} Q' \xrightarrow{*} Q''$, such that $(P', Q'') \in WGB$, according to the second rule of the above construction, $(P', Q') \in WGB'$ ;

1.2 Or $(P,Q) \notin WGB$, then (P, Q) must be produced by the second rule of the above construction. According to the definition of WGB, there must be $S' \xrightarrow{*\alpha} Q' \xrightarrow{*} S''$, such that $(P', Q'') \in WGB$. This shows that $Q \xrightarrow{*} S' \xrightarrow{*\alpha} Q'$, which implies $Q \xrightarrow{*\alpha} Q'$. The reason is simple: the composition of two stars is again a star. Thus it is $(P', Q') \in WGB'$.

2. if $P \xrightarrow{\tau} P'$ then

2.1 Either $(P,Q) \in WGB$, then there is $Q'$ such that $Q \xrightarrow{*} Q'$ and $(P', Q') \in WGB$ which implies $(P', Q') \in WGB'$

2.2 Or $(P,Q) \notin WGB$, then (P, Q) must be produced by the second rule of the above construction. According to the definition of WGB, there must be $S' \xrightarrow{*} Q'$, such that $(P', Q') \in WGB$, which implies $Q \xrightarrow{*} S' \xrightarrow{*} Q'$, or $Q \xrightarrow{*} Q'$. Thus it is $(P', Q') \in WGB'$.

We complete the proof by repeating the above reasoning inductively.

□

The idea of this lemma is simple because we notice that in both transition rules of definition 5.4 the transition that Q performs is either * or $*\alpha*$. Both transitions start with a star *. This tells us that for each $*\alpha*$ we can always move the second star of $*\alpha*$ to its next transition while not affecting its semantics.

Based on this lemma, we will replace the transition $Q \xrightarrow{*\alpha*} Q'$ with $Q \xrightarrow{*\alpha} Q'$ in all definitions about weak bisimularity in the remaining part of this section.

**Lemma 5.2:** Assume name z is defined, then for any q, p, tr, and c,

$\overline{tinput}(tconf\,(q, p, tr), [c], [z])$ is defined if and only if

$\overline{t-barb-i}(tconf\,(q, p, tr), [c])$ is defined.

Proof: Roughly speaking, the axioms of tinput show that the first non-hidden closure is defined if after a finite sequence of $\tau$ actions an input action with c as channel and z as inputted name is feasible. This implies that after a finite sequence of $\tau$ actions an input barb check with c as channel is defined. This implies again that the second non-hidden closure is defined. By reversing the reasoning process the lemma is proved.

□

**Theorem 5.4:** A pair of processes (P, Q) are weakly ground bisimilar iff their corresponding seed algebras A (WGB', a) and A (WGB', b) are deep isomorphic, where a = tconf (p, p, ()), b = tconf (q, q, ()). Note that p and q are the seed algebra element form of the processes P and Q.

Proof: We go the way nhc (A (WGB', a)) $\xrightarrow{1}$ active action (P) $\xrightarrow{2}$ passive action (Q) $\xrightarrow{3}$ nhc (A (WGB', b)) to prove the Only If part of rough seed isomorphism. Consider table 5.3 to be convinced that this conclusion is true.

Note that in table 5.3 we use the notation $tinput'$ to denote a limited version of the non-hidden function $tinput$ of the axiom system of WGB, which obeys the axiom input (i-proc (c, a, p), c, a) = p to meet the requirement of inputting a single fresh name in the definition of weak ground bisimulation. This implies the limitation that any input statement input (i-proc (c, a, p), c, b) = p, where a $\neq$ b, is not considered in weak ground bisimulation. This is different from the non-hidden function $tinput$ used in the axiom system of data type WEB (will be discussed next).

For proving the If part, it is easy to reverse the proof chain to active action (P) $\xrightarrow{1}$ nhc (A (WGB', a)) $\xrightarrow{2}$ nhc (A (WGB', b)) $\xrightarrow{3}$ passive action (Q).

The idea of proving the deep isomorphism is then similar to the third step of the proof of theorem 5.1. We will not repeat it at this place.

□

Table 5.3 rough seed isomorphic mapping for WGB'

| nhc (A (WGB', a)) | Active action | Passive action | nhc (A (WGB', b)) |
|---|---|---|---|
| This place inherits everything from table 5.1 | | | |
| Locating sequence + $\overline{tinput'}$ | Input a single fresh name via a channel | Input the same single fresh name via the same channel | Locating sequence + $\overline{tinput'}$ |
| | | $\tau$ series + Input the same single fresh name via the same channel | h-or-no series + Locating sequence + $\overline{tinput'}$ |
| Locating sequence + $\overline{toutput}$ | Output any name via a channel | Output the same name via the same channel | Locating sequence + $\overline{toutput}$ |
| | | $\tau$ series + Output the same name via the same channel | h-or-no series + Locating sequence + $\overline{toutput}$ |

**Theorem 5.5: F**rom the deep isomorphism of A (WGB’, a) and A (WGB’, b) it follows the deep isomorphism of A (WBB, a) and A (WBB, b), but not vice versa, where a, b have the same meaning as in theorem 5.4.

Proof: The table inherits everything from table 5.1 and rejects nothing. Thus the proof of this theorem is already implied in the proof of theorem 5.4

□

**Example 5.4: (P =** $\tau.x!a$ **, Q =** $x!a$ **)** are weak ground bisimular.

Proof:

p = tau (o-proc (x, a, nil)), q = o-proc (x, a, nil)

Their seed algebras are:

A (WGB, tconf (p, p, ())), A (WGB, tconf (q, q, ()))

We prove that these two seed algebras are really deep isomorphic.

[tconf (p, p, ()))] = {tconf (p, p, ()), h-or-no (tconf (p, p, ())), ……}
= { tconf (p, p, ()), tconf (o-proc (x, a, nil), p, ()), ……};

[tconf (q, q, ()))] = {tconf (q, q, ()), h-or-no (tconf (q, q, ())),……}
= { tconf (q, q, ()), tconf (o-proc (x, a, nil), p, ()), ……};

The non-hidden closures of the seed algebras are nhc (A (WGB, tconf (p, p, ()))) = { $\overline{t-or-no}$ ( tconf (p, p, ())), $\overline{toutput}$ ( tconf (p, p, ()), x, a), $\overline{toutput}$ ( $\overline{t-or-no}$ ( tconf (p, p, ()))), x, a)} nhc (A (WGB, tconf (q, q, ()))) = { $\overline{t-or-no}$ ( tconf (q, q, ())), $\overline{toutput}$ ( tconf (q, q, ()), x, a), $\overline{toutput}\,(\overline{t-or-no}$ ( tconf (q, q, ()))), x, a)}

The rough seed isomorphism between the two non-hidden closures is obvious. We only have to check the deep isomorphism. Since

$\overline{t-or-no}$ ( tconf (p, p, ())) = {tconf (o-proc (x, a, nil), p, ())}

$\overline{t-or-no}$ ( tconf (q, q, ())) = {tconf (o-proc (x, a, nil), q, ())}

is the unique corresponding pair, the deep isomorphism is guaranteed.

□

Now we turn to weak early bisimulation.

**Definition 5.5 (Weak Early Bisimulation)** A symmetric relation WEB on the set of processes is called a weak early bisimulation if for any (P, Q) ∈ WEB,

3.1 For any $\alpha \neq \tau$, $P \xrightarrow{\alpha} P'$, there is a Q', such that $Q \xrightarrow{*\alpha} Q'$ and (P', Q') ∈ WEB;

3.2 For any $P \xrightarrow{\tau} P'$, there is a Q', such that $Q \xrightarrow{*} Q'$ and (P', Q') ∈ WEB.

The difference between early bisimulation and ground bisimulation is the input action. For two processes to be weakly early bisimilar, each non-$\tau$ transition of one process must be matched by a similar transition of the other one, possibly accompanied by a $\tau$ series. This holds in particular for transitions labeled by free-input action. But for weak ground bisimulation, it suffices to consider the simple fresh-name-input. For instance, to establish that $(c?x.P, c?x.Q) \in WGB$, it suffices to show that $(P, Q) \in WGB$, but to establish $(c?x.P, c?x.Q) \in WEB$, it is necessary to show that for every y

$(P[y/x], Q[y/x]) \in WEB$.

The following ADT is for weak early bisimulation.

**Type** WEB =

{ **Type** WGB'

+ **axiom**

input (i-proc(c, x, p), c, y) = subst (p, x, y)

}

**End of Type** WEB

**Theorem 5.6:** Two processes P and Q are weakly early bisimular if and only if the seed algebras A (D, tconf (p, p, ())) and A (D, tconf (q, q, ())) are deep isomorphic, where D = WEB.

Proof: Similar to theorems 5.1, 5.2 and 5.4, where the proofs are delineated with tables, the proof of this theorem is basically illustrated in table 5.4, where one can see how the rough seed isomorphism is constructed. Given the uniqueness of the mapping nhc (A (WEB, a)) $\xrightarrow{1}$ active action (P) $\xrightarrow{2}$ passive action (Q) $\xrightarrow{3}$ nhc (A (WEB, b)) is proved, the proof of rough seed isomorphism and deep isomorphism is straightforward.

□

Table 5.4 rough seed isomorphic mapping for WEB

| nhc (A (WEB, a)) | Active action | Passive action | nhc (A (WEB, b)) |
|---|---|---|---|
| This place inherits everything from table 5.3 | | | |
| Locating sequence + $\overline{tinput}$ | Input any name via a channel | Input the same name via the same channel | Locating sequence + $\overline{tinput}$ |
| | | $\tau$ series + Input the same name via the same channel | h-or-no series + Locating sequence + $\overline{tinput}$ |
| Locating sequence + $\overline{toutput}$ | Output any name via a channel | Output the same name via the same channel | Locating sequence + $\overline{toutput}$ |
| | | $\tau$ series + Output the same name via the same channel | h-or-no series + Locating sequence + $\overline{toutput}$ |

**Theorem 5.7: F**rom the deep isomorphism of A (WEB, a) and A (WEB, b) it follows the deep isomorphism of A (WGB', a) and A (WGB', b), but not vice versa, where a, b have the same meaning as in theorem 5.6.

Proof: Note that table 5.4 inherits everything from table 5.3 and rejects nothing. The proof of this theorem is already implied in the proof of theorem 5.6

□

All transition rules given above are early transition rules. They represent the early semantics of process algebra. Now we will consider late semantics. It is almost the same as early semantics but the input and communication rules.

**Definition 5.6:** The input rule and the communication rule of late semantics are:

Input (Late): $c?u.P \xrightarrow{c(u)}_l P$

Compo (Late): $$\frac{P \xrightarrow{c(u)}_l P', u \notin fn(Q)}{P \mid Q \xrightarrow{c(u)}_l P' \mid Q}$$

Commu (Late): $\dfrac{P \xrightarrow{c(u)}_l P', Q \xrightarrow{c!a}_l Q'}{P|Q \xrightarrow{\tau}_l P'[a/u]|Q'}$

Thus in the late input rule, the instantiation of the placeholder u does not happen in this rule itself. The process P just prepares a placeholder u and 'waits' for any input. The instantiation of the placeholder only happens when the communication happens, as the late communication rule shows. This reminds us of two things. First, the late input rule is semantically separated in two steps. In the first step it prepares for receiving any input. In the second step it receives the inputted variable and completed the instantiation. Second, if no input really happens (i.e. the second step does not happen), then the input rule inputs nothing. The corresponding process just finishes a 'wait for input' action. Therefore we need two non-hidden functions tinput1 and tinput2 to implement the semantics of the two step late input transition. On the other hand, the late communication rule remains the same as the early one.

Now we consider the weak late bisimulation.

**Definition 5.7:** A symmetric relation WLB defined on processes is called weak late bisimulation if for any (P, Q) $\in$ WLB and $P \xrightarrow{\alpha}_l P'$, where bn ($\alpha$) is flash, such that

Case 1: $\alpha = c(u)$, then $\exists Q'$, $Q \xrightarrow{*\alpha}_l Q'$ and $\forall a:$ (P'[a/u], Q' [a/u]) $\in$ WLB;

Case 2: $\alpha$ is not an input, nor a $\tau$ action, then $\exists Q'$, $Q \xrightarrow{*\alpha}_l Q'$ and (P', Q') $\in$ WLB;

Case 3: $\alpha$ is a $\tau$ action, then $\exists Q'$, $Q \xrightarrow{*}_l Q'$ and (P', Q') $\in$ WLB;

**Type** WLB =
{ **Type** WEB +
  **op**
  tinput1: **ptrace** $\times$ **chan** $\times$ **var** → **ptrace**
  tinput2: **ptrace** $\times$ **chan** $\times$ **var** → **ptrace**
  **hop**
  tconf1: **proc** $\times$ **proc** $\times$ **trace** → **ptrace**
  input1: **proc** $\times$ **chan** $\times$ **var** → **proc**
**axiom**
  tinput1 (tconf (q, p, tr), c, u) =
    tconf1 (input1 (q, c, u), p, run (tr, it (c, u)))
  tinput2 (tconf1 (q, p, run (tr, it (c, u))), c, a) =
    tconf (subst (q, u, a), p, run (tr, it (c, a)))
  if free (y, a) = false then input1 (compol (x, y), c, u) = compol (input1 (x, c, u), y)
  if free (x, a) = false then input1 (compor (x, y), c, u) = compor (x, input1 (y, c, u))
  input1 (i-proc (c, u, p), c, u) = p
  }
**End of Type** WLB

**Theorem 5.8:** Two processes P and Q are weakly late bisimular if and only if the seed algebras A (WLB, a) and A (WLB, b) are deep isomorphic, where a = tconf (p, p, ()), b = tconf (q, q, ()).

Proof: With table 5.5, the proof idea is similar to the theorems 5.1, 5.2, 5.4 and 5.6. We get an ADT for WLB by replacing all non-hidden operations and axioms involving tinput with tinput1 and tinput2; all hidden operations and axioms involving input with input1 and subst. The tinput1 function corresponds to the first step of the two step procedure of late input semantics mentioned above, while the tinput2 function corresponds to its second step. Since tinput2 is only allowed to happen after tinput1, the tinput1 axiom does not transform tconf element to tconf element as usually, but transforms tconf element to tconf1 element, where tconf1 is a new function, such that tinput2 is only applicable to tconf1, thus avoiding the application of tinput2 to any other tconf element to invoke a chaos. In case we do not want the input value instantiation to happen, we just need to let the input value ***a*** in the tinput2 axiom equal to the channel name u. Then the only impact of applying tinput2 to the tconf1 element is to turn it back to a tconf element.

□

Table 5.5 rough seed isomorphic mapping for WLB

<table>
<tr><th>nhc (A (WLB, a))</th><th>Active action</th><th>Passive action</th><th>nhc (A (WLB, b))</th></tr>
<tr><td colspan="4">This place inherits everything from table 5.4</td></tr>
<tr><td rowspan="2">Locating sequence + $\overline{tinput1}$</td><td rowspan="2">Prepare an input channel</td><td>Prepare the same input channel</td><td>Locating sequence + $\overline{tinput1}$</td></tr>
<tr><td>$\tau$ series + Prepare the same input channel</td><td>h-or-no series + Locating sequence + $\overline{tinput1}$</td></tr>
<tr><td rowspan="2">$\overline{tinput2}$</td><td rowspan="2">Perform a substitution</td><td>Perform the same substitution</td><td>$\overline{tinput2}$</td></tr>
<tr><td>$\tau$ series + Perform the same substitution</td><td>h-or-no series + $\overline{tinput2}$</td></tr>
</table>

**Theorem 5.9: F**rom the deep isomorphism of A (WLB, a) and A (WLB, b) it follows the deep isomorphism of A (WEB, a) and A (WEB, b), but not vice versa, where a, b have the same meaning as in theorem 5.8.

Proof: Note that table 5.5 inherits everything from table 5.4 and rejects nothing. The proof of this theorem is already implied in the proof of theorem 5.8.

□

Finally, we come to the most powerful bisimulation, the open bisimulation. We first discuss its weak semantics.

**Definition 5.8 (Weak Open Bisimulation)** A symmetric binary relation WOB is called a weak open bisimulation if for each $(P,Q) \in S$ and for all substitutions $\sigma$

(1) if $P\sigma \xrightarrow{\alpha} P'$ where $\alpha$ is $c!x$ or $c?x$, there is $Q'$ such that $Q\sigma \xrightarrow{*\alpha} Q'$ and $(P',Q') \in S$.

(2) if $P \xrightarrow{\tau} P'$ then there is $Q'$ such that $Q \xrightarrow{*} Q'$ and $(P',Q') \in S$.

P and Q are called weakly open bisimilar, if there is a weak open bisimulation WOB such that $(P,Q) \in WOB$.

Open bisimulation is more powerful than late bisimulation. In open bisimulation there is no need for special treatment of input actions since the quantification over substitutions recurs anyway when further transitions are examined: if $P \xrightarrow{c(x)} P'$ is simulated by $Q \xrightarrow{c(x)} Q'$, then the requirement implies that all transitions from $P'\sigma$ must be simulated by $Q'\sigma$ for all $\sigma$, including those substitutions that instantiate x to another name. This also demonstrates the major difference between open bisimulation and late bisimulation: late bisimulation requires the processes to continue to bisimulate under all substitutions for the bound input object, open bisimulation requires them to bisimulate under all substitutions—not only those affecting the bound input object.

So with just a little modification to the ADT WLB, we can get the following ADT for weak open bisimulation.

**Type** WOB =

{ **Type** WLB

+ **op:**

tsubst: **ptrace → ptrace**

+ **axiom**

tsubst (tconf (p, tp, tr)) = tconf (subst (p, x, y), tp, tr)

}

**End of Type** WOB

**Theorem 5.10:** A pair of processes (P, Q) are weakly open bisimilar iff their corresponding seed algebras A (WOB, a) and A (WOB, b) are deep isomorphic, where a = tconf (p, p, ()), b = tconf (q, q, ()). Note that p and q are the seed algebra element form of the processes P and Q.

Proof: With table 5.6, the proof idea is similar to the theorems 5.1, 5.2, 5.4, 5.6 and 5.8. The key is to introduce a non-hidden function tsubst, such that before application of any input or output function we can first apply the tsubst function to implement the principle of open bisimulation: first substitution then transition. Since any composition of substitutions is again a substitution, it is harmless to apply an arbitrary number of tsubst function to a process.

□

Table 5.6 rough seed isomorphic mapping for WOB

| nhc (A (WOB, a)) | Active action | Passive action | nhc (A (WOB, b)) |
|---|---|---|---|
| This place inherits everything from table 5.5 | | | |
| $\overline{tsubst}$ | Perform a substitution | Perform the same substitution | $\overline{tsubst}$ |
| | | $\tau$ series + Perform the same substitution | h-or-no series + $\overline{tsubst}$ |

**Theorem 5.11: F**rom the deep isomorphism of A (WOB, a) and A (WOB, b) it follows the deep isomorphism of A (WLB, a) and A (WLB, b), but not vice versa, where a, b have the same meaning as in theorem 5.10.

Proof: Note that table 5.6 inherits everything from table 5.5 and rejects nothing. The proof of this theorem is already implied in the proof of theorem 5.10.

□

**Definition** 5.9 **(Strong Ground Bisimulation)** A symmetric binary relation SGB is called a strong ground bisimulation if for each $(P,Q)\in SGB$, there is $z\notin fn(P,Q)$ such that if $P\xrightarrow{\alpha}P'$ where $\alpha$ is $c!x$ or $cz$ or $\tau$, there is $Q'$ such that $Q\xrightarrow{\alpha}Q'$ and $(P',Q')\in S$.

P and Q are called strongly ground bisimular, if there is a strong ground bisimulation SGB such that (P, Q)$\in$SGB.

Now we introduce the data type SGB for strong ground bisimulation.

**Type** SGB =

{ **Type** WGB'

**+ op**

t-h-act: **ptrace → ptrace**

**- hop**

h-or-no

**+ axiom**

t-h-act (tconf (q, p, tr)) = tconf (h-act (q), p, tr)

}

**End of Type** SGB

**Theorem 5.12:** A pair of processes (P, Q) are strong ground bisimilar iff their corresponding seed algebras A (SGB, a) and A (SGB, b) are deep isomorphic, where a = tconf (p, p, ()), b = tconf (q, q, ()). Note that p and q are the seed algebra element form of the processes P and Q.

Proof: With table 5.7, the proof idea is similar to the theorems 5.1, 5.2, 5.4, 5.6, 5.8 and 5.10.

□

Table 5.7 rough seed isomorphic mapping for SGB

| nhc (A (SGB, a)) | Active action | Passive action | nhc (A (SGB, b)) |
|---|---|---|---|
| This place inherits everything from table 5.3 | | | |
| Locating sequence + $\overline{tinput}'$ | Input a single fresh name via a channel | Input the same single fresh name via the same channel | Locating sequence + $\overline{tinput}'$ |
| Locating sequence + $\overline{toutput}$ | Output any name via a channel | Output the same name via the same channel | Locating sequence + $\overline{toutput}$ |
| Locating sequence + $\overline{t-h-act}$ | $\tau$ action | $\tau$ action | Locating sequence + $\overline{t-h-act}$ |

**Theorem 5.13: F**rom the deep isomorphism of A (SGB, a) and A (SGB, b) it follows the deep isomorphism of A (WGB, a) and A (WGB, b), but not vice versa, where a, b have the same meaning as in theorem 5.12.

Proof: Note that table 5.7 inherits everything from table 5.2 except the hidden h-or-no function. This can be easily seen when we note that the table 5.2 is part of the table 5.7, with the semantics of its non-hidden closures, its active and passive actions reduced to not include any h-or-no series ($\tau$ series). This is necessary because of the requirement of strong bisimularity. Therefore the conclusion of this theorem is already implied by theorem 5.12. The proof of this theorem is already implied in the proof of theorem 5.12.

□

We know that strong ground bisimulation is not only a fortified version of weak ground bisimulation, but also a fortified version of strong barbed bisimulation. We are interested in seeing whether the data type SGB can also be constructed on the basis of SBB. This is really the case. For the moment we use the notation SGB' to denote the new ADT. While doing that, it was interesting to find that the way from SBB to SGB' is completely parallel to the way from WBB to WGB', such that we can simply write:

**Type** SGB' – **Type** SBB = **Type** WGB' – **Type** WBB

Or, we can also write:

**Type** SGB' = **Type** WGB' + **Type** SBB – **Type** WBB
= **Type** WGB' + **op** t-h-act – **hop** h-or-no
+ **axiom** t-h-act (tconf (q, p, tr)) = tconf (h-act (q), p, tr)

= **Type** SGB

Thus we have the following proposition and theorem:

**Proposition** 5.1: **Type** SGB and **Type** SGB' are the same.

□

**Theorem 5.14: F**rom the deep isomorphism of A (SGB, a) and A (SGB, b) it follows the deep isomorphism of A (SBB, a) and A (SBB, b), but not vice versa, where a, b have the same meaning as in theorem 5.12.

Proof: By using proposition 5.1.

□

**Definition 5.10 (Strong Early Bisimulation)** A symmetric binary relation SEB is called a strong early bisimulation if for each $(P, Q) \in SEB$, if $P \xrightarrow{\alpha} P'$ where $\alpha$ is $c!x$ or $c?x$ or $\tau$ , there is $Q'$ such that $Q \xrightarrow{\alpha} Q'$ and $(P', Q') \in SEB$ .

P and Q are called strong early bisimular, if there is a strong early bisimulation SEB such that (P, Q) $\in$ SEB.

Note that we have used the notation SBS to denote the same strong early bisimulation in section 4. Here we use the notation SEB to show the difference that SEB is introduced based on the underlying type SGB.

**Type** SEB =

{ **Type** WEB

**+ op**

t-h-act: **ptrace → ptrace**

**- hop**

h-or-no

**+ axiom**

t-h-act (tconf (q, p, tr)) = tconf (h-act (q), p, tr)

}

**End of Type** SEB

**Theorem 5.15:** A pair of processes (P, Q) are strong early bisimilar iff their corresponding seed algebras A (SEB, a) and A (SEB, b) are deep isomorphic, where a = tconf (p, p, ()), b = tconf (q, q, ()). Note that p and q are the seed algebra element form of the processes P and Q.

Proof: With table 5.8, the proof idea is similar to the theorems 5.1, 5.2, 5.4, 5.6, 5.8, 5.10 and 5.12.

□

Table 5.8 rough seed isomorphic mapping for SEB

| nhc (A (SEB, a)) | Active action | Passive action | nhc (A (SEB, b)) |
|---|---|---|---|
| This place inherits everything from table 5.4 | | | |
| Locating sequence + $\overline{tinput}$ | Input a single fresh name via a channel | Input the same single fresh name via the same channel | Locating sequence + $\overline{tinput}$ |
| Locating sequence + $\overline{toutput}$ | Output any name via a channel | Output the same name via the same channel | Locating sequence + $\overline{toutput}$ |
| Locating sequence + $\overline{t-h-act}$ | $\tau$ action | $\tau$ action | Locating sequence + $\overline{t-h-act}$ |

**Theorem 5.16: F**rom the deep isomorphism of A (SEB, a) and A (SEB, b) it follows the deep isomorphism of A (WEB, a) and A (WEB, b), but not vice versa, where a, b have the same meaning as in theorem 5.15.

Proof: Here we apply the same reasoning as in theorem 5.13. The proof of this theorem is implied in the proof of theorem 5.15.

□

We are interested in seeing whether the data type SEB can also be constructed on the basis of SGB. We call it SEB’. Like the case of SGB and SGB’, it is easy to see the following:

**Proposition** 5.2: **Type** SEB and **Type** SEB’ are the same.

□

**Theorem 5.17: F**rom the deep isomorphism of A (SEB, a) and A (SEB, b) it follows the deep isomorphism of A (SGB, a) and A (SGB, b), but not vice versa, where a, b have the same meaning as in theorem 5.15.

Proof: By using proposition 5.2.

□

**Definition 5.11 (Strong Late Bisimulation):** A symmetric relation SLB defined on processes is called strong late bisimulation if for any (P, Q) $\in$ SLB and $P \xrightarrow{\alpha}_l P'$, where bn ($\alpha$) is flash, such that

Case 1: $\alpha = c(u)$, then $\exists Q'$, $Q \xrightarrow{\alpha}_l Q'$ and $\forall a$, (P’[a/u], Q’[a/u]) $\in$ SLB;

Case 2: $\alpha$ is $c!x$ or $\tau$, then $\exists Q'$, such that $Q \xrightarrow{\alpha}_l Q'$ and (P’, Q’) $\in$ WLB;

P and Q are called strongly late bisimilar, if there is a strong late bisimulation SLB such that (P, Q) $\in$ SLB.

The data type proposed for strong late bisimularity is the following:

**Type** SLB =

{ **Type** WLB

**+ op**

t-h-act: **ptrace → ptrace**

**- hop**

h-or-no

**+ axiom**

t-h-act (tconf (q, p, tr)) = tconf (h-act (q), p, tr)

}

**End of Type** SLB

**Theorem 5.18:** A pair of processes (P, Q) are strong late bisimilar iff their corresponding seed algebras A (SLB, a) and A (SLB, b) are deep isomorphic, where a = tconf (p, p, ()), b = tconf (q, q, ()). Note that p and q are the seed algebra element form of the processes P and Q.

Proof: With table 5.9, the proof idea is similar to the theorems 5.1, 5.2, 5.4, 5.6, 5.8, 5.10, 5.12 and 5.15.

□

Table 5.9 rough seed isomorphic mapping for SLB

| nhc (A (SLB, a)) | Active action | Passive action | nhc (A (SLB, b)) |
|---|---|---|---|
| This place inherits everything from table 5.5 | | | |
| Locating sequence + $\overline{tinput1}$ | Prepare an input channel | Prepare the same input channel | Locating sequence + $\overline{tinput1}$ |
| $\overline{tinput2}$ | Perform a substitution | Perform the same substitution | $\overline{tinput2}$ |
| Locating sequence + $\overline{t-h-act}$ | $\tau$ action | $\tau$ action | Locating sequence + $\overline{t-h-act}$ |

**Theorem 5.19: F**rom the deep isomorphism of A (SLB, a) and A (SLB, b) it follows the deep isomorphism of A (WLB, a) and A (WLB, b), but not vice versa, where a, b have the same meaning as in theorem 5.18.

Proof: Here we apply the same reasoning as in theorem 5.16. The proof of this theorem is implied in the proof of theorem 5.18.

□

We are interested in seeing whether the data type SLB can also be constructed on the basis of SEB. We call it SLB'. Like the case of SGB and SGB', it is easy to construct SLB' and to see the following:

**Proposition** 5.3: **Type** SLB and **Type** SLB' are the same.

□

**Theorem 5.20: F**rom the deep isomorphism of A (SLB, a) and A (SLB, b) it follows the deep isomorphism of A (SEB, a) and A (SEB, b), but not vice versa, where a, b have the same meaning as in theorem 5.18.

Proof: By using proposition 5.3.

□

**Definition 5.12 (Strong Open Bisimulation):** A symmetric binary relation SOB is called a strong open bisimulation if for each $(P,Q) \in SOB$ and for all substitutions $\sigma$ if $P\sigma \xrightarrow{\alpha} P'$ where $\alpha$ is $c!x$ or $c?x$ or $\tau$, there is $Q'$ such that $Q\sigma \xrightarrow{\alpha} Q'$ and $(P',Q') \in SOB$.

P and Q are called strong open bisimilar, if there is a strong open bisimulation SOB such that $(P,Q) \in SOB$.

Following is the data type SOB we propose for strong open bisimulation.

**Type** SOB =

{ **Type** WOB

**+ op**

t-h-act: **ptrace → ptrace**

**- hop**

h-or-no

**+ axiom**

t-h-act (tconf (q, p, tr)) = tconf (h-act (q), p, tr)

}

**End of Type** SLB

**Theorem 5.21:** A pair of processes (P, Q) are strong open bisimilar iff their corresponding seed algebras A (SOB, a) and A (SOB, b) are deep isomorphic, where a = tconf (p, p, ()), b = tconf (q, q, ()). Note that p and q are the seed algebra element form of the processes P and Q.

Proof: With table 5.10, the proof idea is similar to the theorems 5.1, 5.2, 5.4, 5.6, 5.8, 5.10, 5.12 5.15 and 5.18.

□

Table 5.10 rough seed isomorphic mapping for SOB

| nhc (A (SOB, a)) | Active action | Passive action | nhc (A (SOB, b)) |
|---|---|---|---|
| This place inherits everything from table 5.6 | | | |
| $\overline{tsubst}$ | Perform a substitution | Perform the same substitution | $\overline{tsubst}$ |
| Locating sequence + $\overline{t-h-act}$ | $\tau$ action | $\tau$ action | Locating sequence + $\overline{t-h-act}$ |

**Theorem 5.22: F**rom the deep isomorphism of A (SOB, a) and A (SOB, b) it follows the deep isomorphism of A (WOB, a) and A (WOB, b), but not vice versa, where a, b have the same meaning as in theorem 5.21.

Proof: Here we apply the same reasoning as in theorem 5.19. The proof of this theorem is implied in the proof of theorem 5.21.

□

We are interested in seeing whether the data type SOB can also be constructed on the basis of SLB. We call it SOB'. Like the case of SGB and SGB', it is easy to construct SOB' and to see the following:

**Proposition** 5.4: **Type** SOB and **Type** SOB' are the same.

□

**Theorem 5.23:** If A (SOB, a) and A (SOB, b) are deep isomorphic, then A (SLB, a) and A (SLB, b) are also deep isomorphis, but not vice versa, where a, b have the same meaning as in theorem 5.21.

Proof: By using proposition 5.4.

□

In summarizing, we have built a relational structure of different bisimulations, as shown in figure 5.1. If some arrow in this structure starts from a bisimulation type A and points to another bisimulation type B, then it means B is more restricted than A. That is, $(P,Q) \in B$ implies $(P,Q) \in A$.

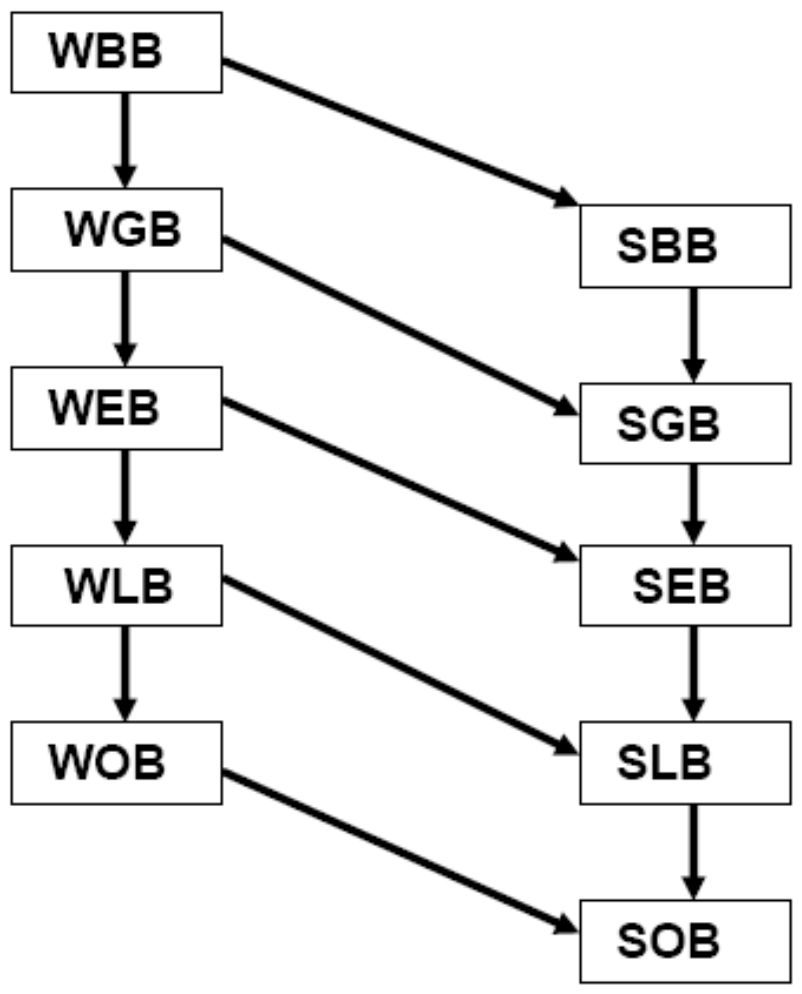

Figure 5.1 A relational picture of different bisimulations

## $ 6 Conclusion

In this paper we introduced an algebraic semantics for process algebra in form of ADT. For that purpose, we developed a particular type of $\Sigma$ algebra, the seed algebra, which describes exactly the behavior of a process. Each non-hidden element of the seed algebra is a state of the process. In the seed algebra, generating an element e1 from another element e2 by the application of a non-hidden algebraic operation represents a state transition from e2 to e1 in process calculus. Therefore a seed algebra can be considered as 'behavior as algebra'. We have shown that two processes bisimulate each other if and only if their corresponding seed algebras have a deep rough seed isomorphism. We have first given a very detailed proof of this theorem for strong early bisimulation. Finally, in section 5, we proved the theorem for all ten different bisimulations and discussed the model theoy of bisimulation with a graph relation between different bisimulation models. We proved that bisimulation B2 subsumes (i.e. is more strict than) bisimulation B1 if and only if the ADT corresponding to B1 is a sub-ADT of that corresponding to B2 in the hieararchy of abstract data types representing the bisimulation family.

Some problems discussed in this paper have been solved in the literature with different techniques. To overcome the difficulty of non-determinism, Broy and Wirsing have assigned each non-deterministic function to its characteristic Boolean valued function[18],while Lescanne proposed to converse non-deterministic functions to functions with set-valued arguments and results[34]. In this paper we introduce instead the concept of non-hidden closures of algebraic elements and their hidden extensions. The isomorphism between sets of non-hidden closures plays a key role in this theory. The algebraic elements contained in a non-hidden closure or a hidden extension are context dependent. This approach of representing nondeterminism is different from those mentioned above.

Our approach also successfully applies the hidden techniques. The hidden property of components has been always an important topic in the study of ADT. Wirsing has given an extensive summary of research results on hidden sorts, hidden functions and hidden signature[47]. The reverse side of hidden is observable. There are lots of works on observability[44, 5]. In the ‘hidden’ approach everything, which is not hidden, is observable, while in the ‘observable’ approach everything, which is not observable, is hidden. It depends on in what the people are interested. Our hidden operation is different from all these works. Our purpose of introducing hidden operation is not to mask the details or private parts of a program module to prevent it from unwanted intrusion. Rather, our purpose is to mask the unwanted details in the isomorphic mapping of seed algebra elements. The introduction of hidden symbols is actually a necessity in establishing the correspondence between seed algebra elements and process states. The granule of masking is also a difference between the various approaches: is it on the signature level[49], the operation level, the sort level[24] or the carrier set elements level[45]? Our approach adopts an operation level for masking.

In summary, this algebraic semantics provides the following advantages:

The first advantage is its character of formality. In the literature, various authors have proposed many different formal semantics for process algebra, including operational semantics (in form of labeled transition systems) and bisimulation semantics (we call it comparative semantics). But all these proposed formal semantics are not as formal as the algebraic semantics. They need more or less the help of natural language to explain the exact meaning they imply. For example, the difference of early and late semantics can hardly be made clear without some literal explanation. But in algebraic semantics, the change of axioms represents everyting rigorously.

The second advantage comes from the multiplicity of its models. The algebraic approach of formal semantics is established as a three level structure: the sorts, the operations and the axioms. Together they form an ADT. That is, the lexicography, the syntax and the semantics are coded separately. One can change any part of this structure while keeping other parts unchanged or basically unchanged. In this way one can obtain different models as different semantics. This shows that the algebraic semantics provides a family of different models. This advantage is shown clearly in section 6. This is also the main advantage of algebraic semantics compared with all other kinds of formal semantics.

The third advantage is its unified form of representation. ADT makes use of an equational logic, which is simple but powerful. Traditionally, syntax and semantics of a process algebra are represented by different formalism. The former is characterized with a BNF like grammer, while the latter with a labeled transition system. In ADT, everything is represented with this three level structure. It is also worthwhile to mention that the context dependency of process algebra, which cannot be represented in form of context free grammar, is representable in our abstract data type formalism.

Our first results of ADT as formal semantics of process algebra were published in [37] as a short report. This paper is a detailed, improved and extended version of [37]. The main new content added in the current version is section 5, the model structure of bisimulation. In [37] we have also announced similar results about quantum process algebra. In order to save space for a in-depth discussion for classical process algebra, we leave out the quantum part in this paper. Our forthcoming paper will be a detailed presentation of results on ADT as formal semantics of quantum process algebra. There is still much more work, which can be done in this direction. First, process algebra, in particular classical process algebra is a very wide field. Many known results in this field can be reconsidered and reexamined with abstract data type as algebraic semantics. At the same time significant new results can be obtained. Second, the algebraic approach presented in this paper can be generalized beyond the area of process algebra, for example it can be applied to other quantum programming languages, which are not process algebra. It would be interesting to compare the theories of qantum programming languages represented in algebraic semantics and represented in other formal semantics.

**Acknowledgement**

We would like to express our sincere thanks to Prof. Mingsheng Ying for his great help by reading through the manuscript of this paper and providing valuable suggestions to improve it. Apart from several technical details, his suggestion includes separating the part of classical process algebra from quantum process algebra, so that the former can be developed more thoroughly in depth and the latter can be developed to form an independent paper, and developing a theory of dominating relations between different types of bisimulation, which now forms the fifth section of this paper.

## Appendix 1 Axiom Part of Process-Trans

**Axiom:**

We use letters p and q to denote variables of sort **proc**, letter x to denote variable of sort **chan**, letters u, v and t to denote variables of sort **name**, string tr to denote variable of sort **trace**,

*#non-hidden operations applied to non-seeds #*

tinput (tconf (p, tp, tr), c, x) = tconf (input (p, c, x), tp, run (tr, it (c, x)))

toutput (tconf (p, tp, tr), c, x) = tconf (output (p, c, x), tp, run (tr, ot (c, x)))

t-h-act (tconf (p, tp, tr)) = tconf (h-act (p), tp, tr)

*#hidden Operations#*

tcall (tconf (p, tp, tr)) = tconf (call (p, tp), tp, tr)

call (compol (x, y), tp) = compol (call (x, tp), y)

call (compor (x, y), tp) = compor (x, call (y, tp))
call (rp, tp) = tp

input (i-proc (c, u, p), c, x) = subst (p, u, x)

output (o-proc (c, x, p), c, x) = p

input (compol(p, q), c, x) = compol(input(p, c, x), q)

input (compor(p, q), c, x) = compor(p, input(q, c, x))

output (compol(p, q), c, x) = compol(output(p, c, x), q)

output (compor(p, q), c, x) = compor(p, output(q, c, x))
h-act (tau (p)) = p
h-act (compo (i-proc (x, u, p), o-proc (x, v, q))) = compo (subst (p, u, v), q)
h-act (compol (p, q)) = compol (h-act (p), q)
h-act (compor (p, q)) = compor (p, h-act (q))

tleft (tconf (sum (p, q), tr)) = tconf (p, tr)
tright (tconf (sum (p, q), tr)) = tconf (q, tr)
tleft (tconf (compo (p, q), tr)) = tconf (compol (p, q), tr)
tright (tconf (compo (p, q), tr)) = tconf (compor (p, q), tr)
tleft (tconf (compol (p, q), tr)) = tconf (compol (left (p), q), tr)
tright (tconf (compol (p, q), tr)) = tconf (compol (right (p), q), tr)
tleft (tconf (compor (p, q), tr)) = tconf (compor (p, left (q)), tr)
tright (tconf (compor (p, q), tr)) = tconf (compor (p, right (q)), tr)
tback (tconf (p, tr)) = tconf (back (p), tr)

left (compo (p, q)) = compol (p, q)
right (compo (p, q)) = compor (p, q)
left (compol (p, q)) = compol (left (p), q)
right (compol (p, q)) = compol (right (p), q)
left (compor (p, q)) = compor (p, left (q))
right (compor (p, q)) = compor (p, right (q))

back (i-proc (x, u, p)) = i-proc (x, u, p)
back (o-proc (x, u, p)) = o-proc (x, u, p)
back (tau (p)) = tau (p)
back (nil) = nil
back (sum (p, q)) = sum (p, q)
back (compo (p, q)) = compo (p, q)
back (compol (p, q)) = compo (back (p), back (q))
back (compor (p, q)) = compo (back (p), back (q))

subst (i-proc (x, u, p), u, t) = i-proc (x, u, p)
subst (o-proc (x, u, p), u, t) = o-proc (x, t, subst (p, u, t))
subst (tau (p), u, t) = tau (subst (p, u, t))
subst (nil, u, t) = nil
subst (sum (p, q), u, t) = sum (subst (p, u, t), subst (q, u, t))
subst (compo (p, q), u, t) = compo (subst (p, u, t), subst (q, u, t))

}
**End of Type** O-process-trans

The last group of axioms describes the invalid compositions of operations, which we will not list here exhaustively.

Appendix 2: Axiom System of WBB

{+ **axiom**

*#Non-hidden operations#*

t-barb-i (tconf (q, p, tr), c) = tconf (barb-i (q, c), p, tr)

t-barb-o (tconf (q, p, tr), c) = tconf (barb-o (q, c), p, tr)

t-or-no (tconf (q, p, tr)) = tconf (or-no (q), p, tr )

*#hidden operation h-or-no #*

h-or-no (tconf (q, p, tr)) = tconf (h-act (q), p, tr)

*#hidden operation h-act #*
h-act (tau (p)) = p
h-act (compo (i-proc (c, u, p), o-proc (c, v, q))) = compo (subst (p, u, v), q)
h-act (compo (o-proc (c, v, q), i-proc (c, u, p))) = compo (q, subst (p, u, v))
h-act (compol (x, y)) = compol (h-act (x), y)
h-act (compor (x, y)) = compor (x, h-act (y))

*#hidden Operation or-no does nothing#*
twrap (tconf (q, p, tr)) = tconf (compo (q, nil), p, tr)

or-no (compo (p, nil)) = compo (p, nil)
or-no (tau (p)) = p
or-no (compo (i-proc (c, u, p), o-proc (c, v, q))) = compo (subst (p, u, v), q)
or-no (compo (o-proc (c, v, q), i-proc (c, u, p))) = compo (q, subst (p, u, v))
or-no (compol (x, y)) = compol (or-no (x), y)
or-no (compor (x, y)) = compor (x, or-no (y))

*# hidden Operation: The locating functions#*
tleft (tconf (q, p, tr)) = tconf (left (q), p, tr)
tright (tconf (q, p, tr)) = tconf (right (q), p, tr)

left (sum (x, y)) = x
right (sum (x, y)) = y

left (compo (x, y)) = compol (x, y)
right (compo (x, y)) = compor (x, y)
left (compol (x, y)) = compol (left (x), y)
right (compol (x, y)) = compol (right (x), y)
left (compor (x, y)) = compor (x, left (y))
right (compor (x, y)) = compor (x, right (y))

*# hidden Operation: recover process state#*
tback (tconf (q, p, tr)) = tconf (back (q), p, tr)

back (i-proc (c, u, p)) = i-proc (c, u, p)
back (o-proc (c, u, p)) = o-proc (c, u, p)
back (tau (x)) = tau (x)
back (nil) = nil
back (sum (x, y)) = sum (x, y)
back (compo (x, y)) = compo (x, y)
back (compol (x, y)) = compo (back (x), back (y))
back (compor (x, y)) = compo (back (x), back (y))

*# hidden Operation: recursive call#*
tcall (tconf (p, tp, tr)) = tconf (call (p, tp), tp, tr)
call (compol (x, y), tp) = compol (call (x, tp), y)
call (compor (x, y), tp) = compor (x, call (y, tp))
call (rp, tp) = tp

*# hidden Operation: check input channel with possible preceding tau actions#*
barb-i (i-proc (c, u, p), c) = i-proc (c, u, p)
barb-i (compol (p, q), c) = compol (barb-i (p, c), q)
barb-i (compor (p, q), c) = compor (p, barb-i (q, c))

*# hidden Operation: check output channel with possible preceding tau actions #*
barb-o (o-proc (c, u, p), c) = o-proc (c, u, p)
barb-o (compol (p, q), c) = compol (barb-o (p, c), q)
barb-o (compor (p, q), c) = compor (p, barb-o (q, c))

*# hidden Operation: substitution#*
subst (i-proc (c, u, p), u, t) = i-proc (c, u, p)
subst (o-proc (c, u, p), u, t) = o-proc (c, t, subst (p, u, t))
subst (tau (x), u, t) = tau (subst (x, u, t))
subst (nil, u, t) = nil

subst (sum (x, y), u, t) = sum (subst (x, u, t), subst (y, u, t))
subst (compo (x, y), u, t) = compo (subst (x, u, t), subst (y, u, t))

}

**End of Type** WBB

Appendix 3 Axiom System of WGB

*#Non-hidden operations#*

tinput (tconf (p, tp, tr), c, x) = tconf (input (p, c, x), tp, run (tr, it (c, x)))

toutput (tconf (p, tp, tr), c, x) = tconf (output (p, c, x), tp, run (tr, ot (c, x)))

*#Hidden Operations#*

input (i-proc (c, x, p), c, x) = p

output (o-proc (c, x, p), c, x) = p

input (compol(p, q), c, x) = compol(input(p, c, x), q)

input (compor(p, q), c, x) = compor(p, input(q, c, x))

output (compol(p, q), c, x) = compol(output(p, c, x), q)

output (compor(p, q), c, x) = compor(p, output(q, c, x))

}

**End of Type** WGB